%% file: 00paper.tex
\documentclass[sigconf,natbib=true,anonymous=false]{acmart}
\graphicspath{{figures/}}

\usepackage[normalem]{ulem}

\AtBeginDocument{%
  \providecommand\BibTeX{{%
    \normalfont B\kern-0.5em{\scshape i\kern-0.25em b}\kern-0.8em\TeX}}}

\copyrightyear{2023} 
\acmYear{2023} 
\setcopyright{acmlicensed}\acmConference[SIGIR '23]{Proceedings of the 46th International ACM SIGIR Conference on Research and Development in Information Retrieval}{July 23--27, 2023}{Taipei, Taiwan}
\acmBooktitle{Proceedings of the 46th International ACM SIGIR Conference on Research and Development in Information Retrieval (SIGIR '23), July 23--27, 2023, Taipei, Taiwan}
\acmPrice{15.00}
\acmDOI{10.1145/3539618.3591883}
\acmISBN{978-1-4503-9408-6/23/07}

\usepackage{booktabs} 
\usepackage{color}
\usepackage{amsfonts}
\usepackage{tabularx}
\usepackage[referable]{threeparttablex}
\usepackage{makecell}
\usepackage{caption}
\usepackage{subcaption}
\usepackage[inline]{enumitem}
\usepackage{multirow}

\newcommand{\eg}{e.g., }
\newcommand{\ie}{i.e., }
\newcommand{\captionshrink}{\vspace{-0.75\baselineskip}}

\begin{document}

\fancyhead{}
\title{MG-ShopDial: A Multi-Goal Conversational Dataset for e-Commerce}

\author{Nolwenn Bernard}
\affiliation{%
  \institution{University of Stavanger}
  \city{Stavanger}
  \country{Norway}
}
\email{nolwenn.m.bernard@uis.no}

\author{Krisztian Balog}
\affiliation{%
  \institution{University of Stavanger}
  \city{Stavanger}
  \country{Norway}
}
\email{krisztian.balog@uis.no}

\begin{abstract}
Conversational systems can be particularly effective in supporting complex information seeking scenarios with evolving information needs.
Finding the right products on an e-commerce platform is one such scenario, where a conversational agent would need to be able to provide search capabilities over the item catalog, understand and make recommendations based on the user's preferences, and answer a range of questions related to items and their usage.
Yet, existing conversational datasets do not fully support the idea of mixing different conversational goals (i.e., search, recommendation, and question answering) and instead focus on a single goal.
To address this, we introduce MG-ShopDial: a dataset of conversations mixing different goals in the domain of e-commerce.
Specifically, we make the following contributions.
First, we develop a coached human-human data collection protocol where each dialogue participant is given a set of instructions, instead of a specific script or answers to choose from.
Second, we implement a data collection tool to facilitate the collection of multi-goal conversations via a web chat interface, using the above protocol.
Third, we create the MG-ShopDial collection, which contains 64 high-quality dialogues with a total of 2,196 utterances for e-commerce scenarios of varying complexity. The dataset is additionally annotated with both intents and goals on the utterance level.
Finally, we present an analysis of this dataset and identify multi-goal conversational patterns.
\end{abstract}



\maketitle

\input{sigir2023-woz-01}
\input{sigir2023-woz-02}

\input{sigir2023-woz-03}

\input{sigir2023-woz-04}
\input{sigir2023-woz-05}
\input{sigir2023-woz-06}
\input{sigir2023-woz-07}

\begin{acks}
We thank all the volunteers and annotators who contributed to the creation of MG-ShopDial.
This research was supported by the Norwegian Research Center for AI Innovation, NorwAI (Research Council of Norway, project number 309834).
\end{acks}

\bibliographystyle{ACM-Reference-Format}
\bibliography{sigir2023-woz.bib}

\end{document}

%% file: sigir2023-woz-01.tex
\section{Introduction}

How many queries would a user need to find a new pair of running shoes? One, if they knew exactly which one they wanted---however, it is not the case for the majority of users. Hence, they would most likely start an information-seeking process to explore the search space and execute several ``context-free'' queries to find the perfect pair of shoes~\citep{RusselRose:2013:book}. For such situations, conversational systems have gained attention, as they offer several advantages over traditional means of information access.
A dialogue is more natural, especially for complex information needs that might require a sequence of (co-dependent) queries. Indeed, the multi-turn structure of conversational interactions allows for making references easily to previous answers, which is not possible in a traditional search engine.
Other advantages include more direct feedback, as it can be expressed in plain text, and better personalization capabilities thanks to the preferences disclosed in the dialogue~\citep{Zamani:2022:arXiv}.
Conversational information access is, in part, supported by the development of natural language processing and deep learning techniques~\citep{Gao:2022:arXiv,Zhang:2020:SciChinaTechSci}, and the growing adoption of conversational assistants~\citep{Voicebot:2022:website}. 
However, advancements in the field are highly dependent on the availability of suitable conversational datasets.

Conversational information access systems support multiple conversational goals that are related to complex information seeking, exploratory information gathering, and recommendation~\citep{Culpepper:2018:SIGIRForum}.
In this work, we focus on the three main conversational goals identified in the field: \emph{search}, \emph{recommendation}, and \emph{question answering (QA)}~\citep{Zamani:2022:arXiv}. The distinction between these goals can sometimes be blurry as the same situation can be considered as search or recommendation (\eg finding a close by hotel), or as search or question answering (\eg agent answering a sequence of questions with passages)~\citep{Zamani:2022:arXiv}. 
Crucially, in a natural conversation, the conversational goal is \emph{dynamic}, \ie it changes depending on the context.
Therefore, a truly conversational information access system should support all these goals.
Taking the example of a user looking to purchase running shoes, the agent starts by eliciting the user's preferences and makes recommendations based on them (\ie \emph{recommendation}). Before making a final choice, the user asks questions related to the eco-friendliness and the sole's characteristics of a suggested pair of shoes (\ie \emph{search} and \emph{QA}).
As the conversation progresses, the conversational goal can be updated several times based on the context. Following the example, the user might not be satisfied by the sole's characteristics of the first recommended pair of shoes and therefore asks for another recommendation.
Hereinafter, we refer to similar conversations that mix goals as \emph{multi-goal conversations}.

Most conversational datasets are created for a particular conversational goal, such as search (\eg CAsT-19~\citep{Dalton:2020:SIGIR} and MISC~\citep{Thomas:2017:CAIR}) or recommendation (\eg MultiWoZ~\citep{Budzianowski:2018:EMNLP} and INSPIRED~\citep{Hayati:2020:EMNLP}).
Few datasets with multi-goal conversations exist, focusing on the domains of music, movie, restaurant, and news~\citep{Dodge:2016:ICLR,Liu:2021:EMNLP,Liu:2020:ACL}. However, they do not support all three goals studied here.
Hence, we create MG-ShopDial, a new dataset of English multi-goal conversations in the domain of e-commerce.
According to~\citet{Papenmeier:2022:CHIIR}, e-commerce could take advantage of the conversational setting, especially for product search. Indeed, clients do not always know with precision what they are looking for, likely resulting in a multi-goal conversation (\eg exploration of the space, disclosure of information need, clarification questions).
Thus, MG-ShopDial is a resource that is particularly suited for the development of future conversational agents that can handle the evolution of a user's conversation goals as naturally as possible.

To collect the conversations, we use a coached human-human data collection protocol~\citep{Radlinski:2019:SIGDIAL}, where some participants mimic digital shopping assistants, while others play the clients.
Each role comes with a set of instructions to detail what is expected with regards to conversational goals. However, we do not suggest answers or provide a specific script to follow. 
Instead, we emphasize on the naturalness of the conversation, the curiosity for the client, and that the shopping assistant should ask clarification questions, if necessary.
In order to introduce diversity in the conversations, two types of scenarios (simple and complex) across different product categories (\eg book, clothes, office supplies) are developed.

The creation of the dataset is facilitated by a purpose-built tool we developed to collect data in accordance with our protocol. Due to poor engagement from crowd workers, data collection is performed with volunteers to ensure that the dialogues are of high quality. As a result, MG-ShopDial contains 64 conversations with a total of 2,196 utterances. 
Upon analyzing MG-ShopDial, we observe a consistent conversational pattern that involves two or three distinct phases: first recommendation, then information seeking, and in some instances, a secondary recommendation. Notably, we find this pattern to be consistent across diverse scenarios and product categories.
When comparing the goals in terms of associated intents, there does not appear to be a clear distinctions between \emph{search} and \emph{QA}, but we do observe a difference between those two goals and \emph{recommendation}.
These observations validate the usefulness of MG-ShopDial for research on multi-goal conversations.

In summary, the main contributions of this work are fourfold. First, we propose a protocol to collect realistic multi-goal conversations in the e-commerce domain. 
Second, we design and implement a tool to perform data collection following our protocol.
Third, we collect and release a dataset of English multi-goal conversations.
Finally, we present a concise analysis of the dataset.
All resources developed as part of this study (MG-ShopDial, data collection tool, and annotation details) are made publicly available at: \url{https://github.com/iai-group/MG-ShopDial}.

%% file: sigir2023-woz-02.tex
\section{Related work}
\label{sec:related}

We present previous work on conversational information access, data collection methodologies, and conversational datasets.

\subsection{Conversational Information Access}
\emph{Conversational information access} systems, also referred to as \emph{conversational information seeking} systems, are defined as agents capable of satisfying information needs through a conversation involving a sequence of interactions~\citep{Zamani:2022:arXiv}. Conversational search, QA, and recommendation are regarded as subdomains of conversational information access, each with their own specificity. 
In the past few years, there has been continuous development on specific subtasks, such as query rewriting~\citep{Lin:2021:TOIS,Yu:2020:SIGIR}, preference elicitation~\citep{Kostric:2021:RecSys,Zuo:2022:CIKM}, answer rewriting~\citep{Baheti:2020:ACL,Szpektor:2020:WWW}, and user intent prediction~\citep{Qu:2019:CHIIR,Cai:2020:UMAP}. 
Some of these subtasks are particular to one of the subdomains mentioned above, like preference elicitation for recommendation and query rewriting for search. Hence, we notice that most studies focus on a single subdomain without considering how all these subtasks could be incorporated into a holistic conversational information access system.
Our work aims to enable progress in that direction.

\subsection{Data Collection Methodologies} 
There are various ways to collect conversational data. One approach is to get logs from an existing system, as was done, for example, for the Carnegie Mellon Communicator Corpus~\citep{Bennett:2002:ICSLP}. 
Another way is to perform user studies focusing on a task, typically using a crowdsourcing platform to recruit participants as in~\citep{Budzianowski:2018:EMNLP,Choi:2018:EMNLP,Zhou:2020:COLING}. 
There are different ways to set up the crowdsourcing task, for example, \citet{Zhou:2020:COLING} ask crowd workers to edit utterances proposed by a neural model, while \citet{Hayati:2020:EMNLP} pair crowd workers to discuss about movies.
The Wizard-of-Oz (WoZ) methodology is a popular approach to collect human-human dialogues. In this setting, there is a human intermediary acting as the conversational agent (i.e., the ``wizard'') and another human interacting with it not knowing that it is a human. Having a human intermediary tackles some practical limitations of conversational systems, especially regarding the understanding of natural language and tracking of conversation state.
Still, some limitations remain, such as the dependence on the task studied (\eg recommendation of hotels vs. question-answering about movies) and the tools available to the wizard~\citep{Serban:2018:DD}. To mitigate this dependence, \citet{Radlinski:2019:SIGDIAL} propose a new methodology derived from WoZ based on the idea of coaching the wizard rather than suggesting answers. For MG-ShopDial, we follow a similar approach, which is described in detail in Section~\ref{sec:protocol}.

There exist platforms to perform dialogue collection for different domains and tasks. \citet{Miller:2017:arXiv} propose the ParlAI platform for dialogue research, with question answering and goal-oriented dialogue among the list of supported tasks, and the possibility to collect dialogues via Amazon Mechanical Turk. 
The CoCoA framework offers dialogue collection tools for three tasks (finding mutual friends, price negotiation, and deal or no deal)~\citep{He:2018:EMNLP,He:2017:ACL}. \citet{Ogawa:2020:LREC} develop a platform based on video games to gamify dialogue collection. Their platform is presented as an alternative to crowdsourcing that has limitations to collect good quality dialogues (\eg workers' motivation, costs). 
Notably, none of these tools support the task studied in this work, \ie multi-goal conversations.  Therefore, we propose our own dialogue collection application: Coached Conversation Collector (CCC), which is described in Section~\ref{sec:ccc}.

\begin{table*}
	\centering
 \captionshrink
	\begin{threeparttable}
		\caption{Comparison of conversational datasets selected. CS: conversational search, CR: conversational recommendation, CQA: conversational question answering, Meta: meta-communication.}
        \captionshrink
		\label{tab:datasets}
		\begin{tabularx}{\linewidth}{ X | c  | X | c | c | c | c | X | X }
			\Xhline{1.2pt}
			Dataset & Language & Participants & CQA & CS & CR & Meta\tnotex{tn:naturalness} & Domains & \# conversations \\ 
			\Xhline{1.2pt}
			Movie dialogue datasets~\citep{Dodge:2016:ICLR} & EN  &  & \checkmark & $\times$ &  \checkmark & \checkmark & Movies & $\sim$120,000 ... $\sim$1M \\ \hline %
			MISC~\citep{Thomas:2017:CAIR} & EN  & Volunteers  & $\times$ & \checkmark &  $\times$ & $\times$ & Open domain & 88 \\ \hline
			QuAC~\citep{Choi:2018:EMNLP} & EN  & Crowd workers & \checkmark & $\times$ &  $\times$ & $\times$ & Open domain & 13,594 \\ \hline
			SQuAD 2.0~\citep{Rajpurkar:2018:ACL} & EN  & Crowd workers & \checkmark & $\times$ &  $\times$ & $\times$ & Open domain & 151,054 (questions) \\  \hline
			MultiWoZ~\citep{Budzianowski:2018:EMNLP} & EN  & Crowd workers in a Wizard-of-Oz set up & $\times$ & $\times$ &  \checkmark & $\times$ &  Restaurants, hotels, attractions, taxis, trains, hospitals, police & 8,438 \\   \hline
			ReDial~\citep{Li:2018:NIPS} & EN & Crowd workers & $\times$ & $\times$ &  \checkmark & \checkmark & Movies & 10,006 \\			\hline
			CAsT-19~\citep{Dalton:2020:SIGIR} & EN  & Experts & $\times$ & \checkmark &  $\times$ & $\times$ & Open domain & 80 \\ \hline
			DoQA~\citep{Campos:2020:ACL} & EN  & Crowd workers in a Wizard-of-Oz set up & \checkmark & $\times$ &  $\times$ & $\times$ & Cooking, travel,  movies & 2,437 \\  \hline
			DuRecDial~\citep{Liu:2020:ACL} & ZH  & Unclear\tnotex{tn:DuRecDial} & \checkmark & $\times$ &  \checkmark & \checkmark & Movies, music, movie stars, food, restaurants, news, weather & 10,190 \\	\hline
			TG-ReDial~\citep{Zhou:2020:COLING} & ZH  & Crowd workers & $\times$ & $\times$ & \checkmark & \checkmark & Movies & 10,000 \\ \hline
			INSPIRED~\citep{Hayati:2020:EMNLP} & EN  & Crowd workers & $\times$ & $\times$ &  \checkmark & \checkmark & Movies & 1,001 \\ \hline
			DuRecDial 2.0~\citep{Liu:2021:EMNLP} & EN, ZH  & Crowd workers & \checkmark & $\times$ &  \checkmark & \checkmark & Movies, music, movie stars, food, restaurants, news, weather & 16,482 \\
			\Xhline{1.2pt}
			MG-ShopDial & EN  & Volunteers &  \checkmark  &   \checkmark &  \checkmark   &  \checkmark  & e-commerce & 64 \\
			\Xhline{1.2pt}
		\end{tabularx}
		\begin{tablenotes}
			\footnotesize
			\item[\textdagger] \label{tn:naturalness} Include sections of a conversation which do not contribute to the completion of the goal but to make the conversation fluid and natural.
            \item[\textdaggerdbl] \label{tn:DuRecDial} Each conversation involves 2 persons, a seeker and a recommender, but their qualification is unclear.
		\end{tablenotes}
	\end{threeparttable}
    \vspace*{0.5\baselineskip}
\end{table*}

\subsection{Conversational Datasets}

Conversational information access has received a growing attention from the information retrieval, dialogue systems, and natural language processing communities. 
Thus, there is a large number of conversational datasets available for various tasks and goals~\citep{Allouch:2021:Sensors,Joko:2021:SIGIR}.\footnote{List of conversational datasets by~\citet{Joko:2021:SIGIR}: \url{https://t.co/4315joogAk?amp=1}}
These datasets can be classified based on the three conversational goals identified: \emph{QA}, \emph{search}, and \emph{recommendation}. 
Note that in the context of conversational information access, we do not consider the social chat problem. Therefore, we analyze datasets of task-oriented and question answering conversations in regards to the above goals. 
To the best of our knowledge, none of the existing datasets contain all three goals. Table~\ref{tab:datasets} compares MG-ShopDial with a selection of well-established and recent datasets.
From the selection of datasets, only the Movie dialogue datasets~\citep{Dodge:2016:ICLR} contain synthetic conversations.
The majority of large human-human datasets use crowdsourcing~\citep{Rajpurkar:2018:ACL,Li:2018:NIPS,Zhou:2020:COLING}, with the Wizard-of-Oz (WoZ) setup employed for some~\citep{Budzianowski:2018:EMNLP,Campos:2020:ACL}. 
Conversely, smaller datasets such as CAsT-19~\citep{Dalton:2020:SIGIR} and MISC~\citep{Thomas:2017:CAIR} are created with volunteers or experts.
Regarding the target domain, we notice that most datasets for \emph{recommendation} focus on movies~\citep{Zhou:2020:COLING,Hayati:2020:EMNLP,Li:2018:NIPS}, while datasets for \emph{QA} and \emph{search} tend to target more than a single domain~\citep{Dalton:2020:SIGIR,Choi:2018:EMNLP,Rajpurkar:2018:ACL}. 
Moreover, few datasets mix conversational goals. It is worth noting that meta-communication is becoming a valuable characteristic and is included in most recent datasets.  
Below, we briefly discuss about existing datasets mixing goals and motivate the need for the development of new ones that target multi-goal conversations.

\begin{table*}
	\centering
	\caption{Checklists for participants.}
    \captionshrink
	\label{tab:checklist}
	\begin{tabularx}{\linewidth}{lX|lX}
		\Xhline{1.2pt}
		\multicolumn{2}{l}{\textbf{Shopping assistant}}  & \multicolumn{2}{l}{\textbf{Client}} \\
		\Xhline{1.2pt}
		(A1) & Greetings & (C1) & Greetings \\ \hline
		(A2) & Determine client's need (e.g., What are they looking for? Do they have any constraints?) & (C2) & Inform the retail assistant about what you are looking for \\ \hline
		(A3) & Ask clarification questions if necessary & (C3) & Inform the retail assistant about your preferences \\ \hline
		(A4) & Elicit client's preferences  & (C4) & Ask factual questions about the recommended items \\ \hline
		(A5) & Make a first recommendation of 3-4 items that fit client's need and preferences & (C5) & Ask general questions about the recommended items \\ \hline
		(A6) & Answer factual questions asked by the user & (C6) & Based on previous utterances ask the retail assistant to refine the list of recommended items with new constraints \\ \hline
		(A7) & Answer general questions asked by the user & (C7) & Choose an item to buy or express your dissatisfaction with the recommended items \\ \hline
		(A8) & Refine the recommendation if the client's need or preferences change & (C8) & Gracefully end the conversation \\ \hline
		(A9) & Gracefully end the conversation & & \\
		\Xhline{1.2pt}
	\end{tabularx}
\end{table*}

\subsubsection{Multi-goal Conversational Datasets}

\citet{Dodge:2016:ICLR} release five datasets that aim to test the abilities of end-to-end dialogue systems. Among these, only two, QA+Recommendation Dataset and Joint Dataset, contain dialogues that mix goals. Yet, the synthetic nature of the dialogues make them unrealistic.
\citet{Liu:2020:ACL} release a first dataset mixing conversational goals, which they call dialogue types. This dataset aims to tackle their newly introduced task: \emph{``conversational recommendation over multi-type dialogues, where the bots can proactively and naturally lead a conversation from a non-recommendation dialogue (e.g., QA) to a recommendation dialogue, taking into account user’s interests and feedback''}~\citep{Liu:2020:ACL}. However, the conversations in this dataset are in Chinese, which limits the scope of research.
To tackle this, \citet{Liu:2021:EMNLP} propose a second version of the dataset including data both in Chinese and English. Furthermore, this second version supports multilingual and cross-lingual conversational recommendation research.

The closest datasets to our problem are \citep{Liu:2020:ACL,Liu:2021:EMNLP}, but unlike us, these do not support conversational search. Also, instead of asking the crowd worker about their actual preferences, they provide a user profile with information such name, gender, occupation, and preferences.
An approach to create a dataset mixing goals is to combine several datasets focusing on one goal is presented in~\citep{Dodge:2016:ICLR}, however, this approach is not considered in this work for the following reasons. First, it would require to have datasets with conversations in the e-commerce domain for the different goals. Second, the conversations would need to be altered in order to integrate the new goals, which would likely lead to a loss of naturalness. This motivates our work to create a new dataset with natural multi-goal conversations.

%% file: sigir2023-woz-03.tex
\section{Protocol}
\label{sec:protocol}

The goal of this paper is to collect multi-goal conversations in the domain of e-commerce using a coached (human-human) data collection protocol. 
For the collection of conversations, we follow the example of \citet{Radlinski:2019:SIGDIAL} by providing general instructions rather than possible answers to choose from. As in~\citep{Radlinski:2019:SIGDIAL}, this method can help to reduce the bias towards to the system, and permit human-level natural language understanding and generation. 
However, we customize our approach to a product search scenario, thereby differing from and extending~\citep{Radlinski:2019:SIGDIAL} in several ways:

\begin{enumerate}[leftmargin=0.5cm]
    \item Instructions are provided to both parties in the conversation: the shopping assistant and the client. Indeed, due to the complexity of the task and to ensure the presence of multiple conversational goals, both clients and assistants need to be coached.
    \item The instructions contain a \emph{checklist} of actions to complete. This checklist allows the participants to easily and quickly assess what remains to be done during the conversation. Table~\ref{tab:checklist} presents the checklists for the shopping assistant and client. To guarantee that the assistant can make a recommendation, they need to uncover the client's need and preferences; this is reflected in actions A3 and A4. Accordingly, the client needs to disclose this information as indicated in actions C2, C3, and C6. The client should also be curious and ask different types of question about the recommended items to explore the search space and make an informed decision (actions C4 and C5).
    \item The conversation has a time limit in order to help participants stay focused and minimize digression.
    \item Conversations are collected for selected product categories. In this work, we select 4 product categories from the Amazon Product dataset~\citep{Ni:2019:EMNLP}: \textsc{Sports and Outdoors}, \textsc{Books}, \textsc{Office Products}, and \textsc{Cell Phones and Accessories}. The motivation behind this choice is that the categories are diverse and include products from daily life and hobbies.
    \item For each category, different scenarios with specific levels of \emph{constraints} and \emph{complexity} are developed. Constraints are divided into 3 levels: low, medium, and high. The low level corresponds to restriction only on the product, \eg \emph{``You are looking for a book in the genre of your choice.''} For the medium level, another constraint is added on top of the product, for example a specific color or budget. Finally, when the scenario  includes at least two constraints in addition to the product to buy, it is classified as a conversation with high constraints, such as \emph{``You are looking for a pair of red running shoes in size 7 [...] made from recycled material.''}
    Moreover, the conversation is considered \emph{simple} when the client is only looking for a unique product, while it is seen as \emph{complex} if more products are sought, \eg a client is looking to buy the necessary equipment to play ice hockey. We refer to the GitHub repository for the detailed specifications of scenarios.\footnote{\url{https://github.com/iai-group/MG-ShopDial/blob/main/CCC/ccc/app/chat/static/yml/topics.yml}}
    \item For each category, we provide a curated \emph{list of products} that can meet the requirements presented in the different scenarios. The shopping assistant has access to the list of products to help them reduce the time needed to generate a response, as too long waiting time could negatively impact the user experience and engagement in the conversation. However, they are allowed to recommend products that are not in the list if they know a better match to the client's need. 
    \item The participants can send text or image utterances. All contemporary e-commerce platforms have images in addition to the textual descriptions of items, thus, we believe that allowing image utterances is more realistic than just text. 
\end{enumerate}

\noindent
Regarding (6), we note that the size or composition of the product list is not a significant factor in this study. Our primary objective is to gain insight into the structure and evolution of conversational goals, rather than to optimize product recommendations among the sea of choices on an e-commerce platform. Consequently, the shopping assistant need not focus on providing the best recommendation from an extensive list of products, but rather on recommending items that align with the client's needs and preferences.

We further note that, different from~\citep{Radlinski:2019:SIGDIAL}, ours is technically not a Wizard-of-Oz protocol, as the client is aware that the role of the shopping assistant is played by another human. This, however, could be changed by adjusting the instructions given to clients.

%% file: sigir2023-woz-04.tex
\section{Coached Conversation Collector}
\label{sec:ccc}

We design and implement an application, Coached Conversation Collector (CCC), to facilitate data collection in accordance with the protocol described in the previous section.
The goal of this application is to match shopping assistants with customers in chat rooms where conversations take place. In addition, the tool allows participants to keep track of their progress using their checklists.
CCC is a modular application that can be adapted to other use cases, by changing the instructions or modifying the chat room interface. 
Below, we describe the main components of CCC (Section~\ref{sec:ccc:components}), followed by implementation details (Section~\ref{sec:ccc:imp}).

\subsection{Components}
\label{sec:ccc:components}

The application is divided into three main components: lobby, chat rooms, and administrator page.
To access the application, users need to register first and specify their assigned role, \ie shopping assistant or client. Shopping assistants also need to specify which categories they are interested in, the first time they log in.

\subsubsection{Lobby}
\begin{figure}
    \vspace*{-\baselineskip}
	\includegraphics[width=\linewidth]{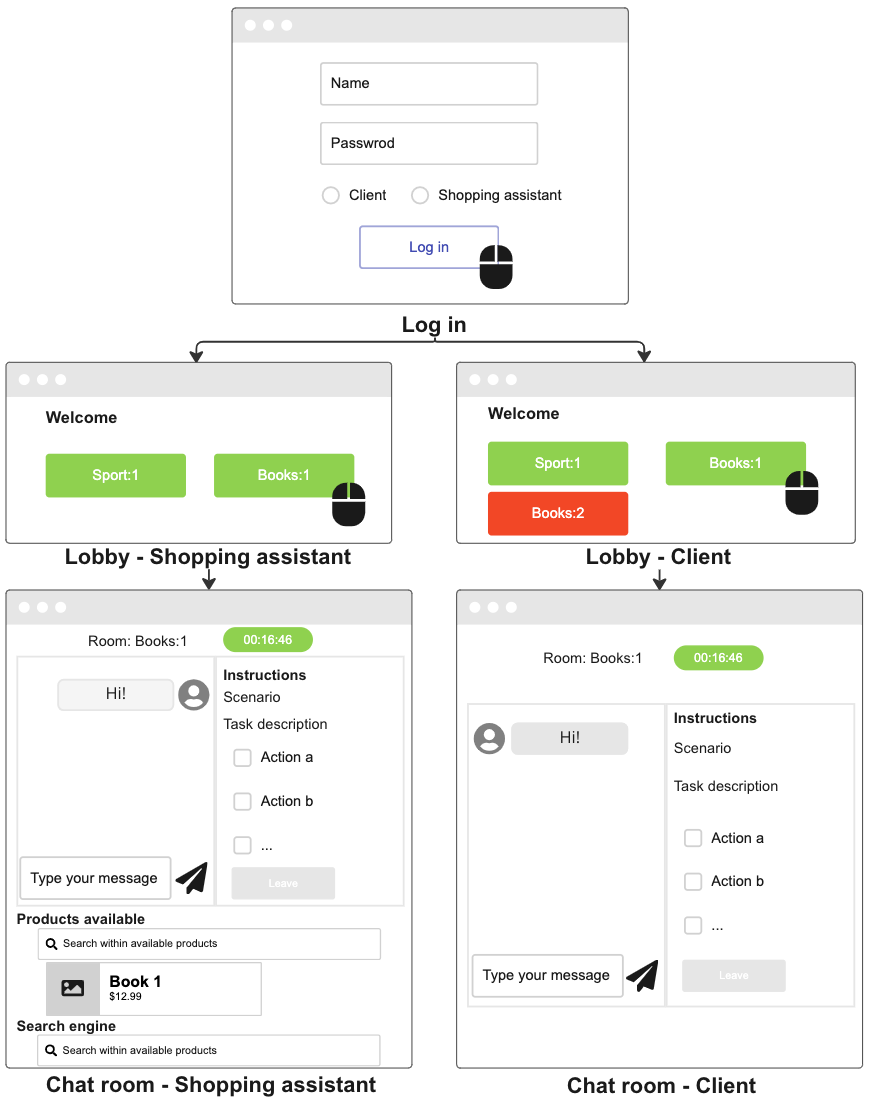}
    \vspace*{-\baselineskip}
	\caption{Mock of the Coached Conversation Collector application, showing the different interfaces per roles.}
	\label{fig:chatroom}
\end{figure}

After logging in, participants are redirected to the lobby. There, they can see the chat rooms available (see the ``Lobby'' mocks in Figure~\ref{fig:chatroom}). 
On the one hand, a shopping assistant only sees their chat rooms that correspond to the categories they selected. For example, if a shopping assistant is interested in \textsc{Sports and Outdoors}, then a room for this category is created and is made available. On the other hand, the client sees a list of all rooms with a color indicating their availability: an occupied room appears in red, while a free room is displayed in green. 

\subsubsection{Chat room}
The interface in a chat room is divided into several elements and differs depending on the role of the participant.
The \emph{client interface} has, on top, a timer displaying the remaining time of the conversation. Below, the main panel is divided into two vertical blocks: the ongoing conversation is displayed on the left, while the instructions related to the task are shown on the right (see ``Chat room - Client'' in Figure~\ref{fig:chatroom}).
In addition to the above elements, the \emph{shopping assistant interface} also includes a product list and access to a search engine (see ``Chat room - Shopping assistant'' in Figure~\ref{fig:chatroom}). The product list is comprised of curated products with descriptions and pictures. 
Moreover, the shopping assistant can search a large web corpus on the paragraph level to answer information needs that they cannot answer with product information or from their own knowledge.
This last element can be used to collect the search logs related to a conversation for further analysis.
The advantage of this interface is its modular aspect. Indeed, the study leader can decide which elements of the interface are needed and can remove the unnecessary ones. For example, the instructions for each participant are defined in a HTML file that is easily editable.

\subsubsection{Administrator page}
The application also has a password protected page where the administrator can see the active users and the chat rooms opened, and access the recorded conversations. In the future, we plan to extend the administration interface with aggregate statistics over the collected conversations (\eg average number of turn per conversation, number of conversations per categories). %

\subsection{Implementation}
\label{sec:ccc:imp}

CCC is implemented in Python, based on the Flask framework\footnote{https://flask.palletsprojects.com/en/2.2.x/} as webserver and using Redis\footnote{https://redis.io} as a message broker and database.
For this work, we index the TREC CAsT 2022 corpora\footnote{\url{https://github.com/daltonj/treccastweb}} (MS MARCO V2 dataset~\citep{Bajaj:2016:arXiv}, KILT Wikipedia~\citep{Petroni:2021:NAACL}, TREC Washington Post 2020\footnote{\url{https://trec.nist.gov/data/wapost/}}) for the internal web search engine. 
The product lists associated to each product category are curated manually from real products available on Amazon.\footnote{\url{https://www.amazon.com}}
Regarding chat messages, participants can send utterances in different modalities: text or image via its URL. 
For each conversation, CCC stores the following metadata, in addition to the timestamped utterances: participants' checklists, search logs, and scenario information. 

%% file: sigir2023-woz-05.tex
\section{Data collection}
\label{sec:collection}

This section presents the data collection procedure (Section~\ref{sec:collection:procedure}) and the participants involved (Section~\ref{sec:collection:participants}), followed by the annotation of conversations in terms of intent and goal (Section~\ref{sec:collection:annotation}). Finally, we give a brief summary of the MG-ShopDial dataset (Section~\ref{sec:collection:sum}).

\subsection{Procedure}
\label{sec:collection:procedure}

The data was collected over the course of several sessions. The sessions were conducted in English, using two formats: in-person with approximately ten participants per session, and remote with a one-on-one format.
At the beginning of each session, the study leader provided an introduction to the task and presented the instructions.

The shopping assistant can choose which category they are interested in (based on their familiarity with the topic) and join the associated chat room. Once the shopping assistant is in the room, the client can join as well.
After joining the chat room, the shopping assistant and the client can read and follow the instructions (\ie description of the task and actions checklist) associated with the task. It is important to note that the shopping assistant does not know beforehand what product the client is looking for, only the product category. This is by design to encourage them to ask questions to uncover the client's need. 

The duration of the conversation takes into consideration the time needed to carefully read the instructions, the latency to get a reply, and the complexity of the task. 
Several experiments were conducted with different duration before settling on 17 minutes as the limit. We tried 13 and 15 minutes, but found those too short, as participants barely had the time to start the information seeking process. A duration set to 20 minutes, however, was slightly too long as participants finished sooner or diverged from the initial scenario. 
During the conversation, all utterances were stored as well as the search logs from the shopping assistant.
It is possible that some conversations contain incorrect information, especially from the shopping assistant (\eg wrong price, bargain campaigns), however it is not an issue for this work, since our interests lie in the discovery of conversational patterns in multi-goal conversations. 

\subsection{Participants}
\label{sec:collection:participants}

Initially, our plan was to conduct the data collection on a crowdsourcing platform.  However, we experienced poor engagement from crowd workers.
For example, some participants left the chat room because they did not get answers quickly enough. Others replied only with very short utterances that did not satisfy the requirements. Also, some workers did not follow the instructions and were chatting on unrelated topics. Therefore, we decided to perform data collection on a smaller scale with the help of volunteers, who were trained to perform the task, in order to collect better quality conversations. 

In total, 21 volunteers participated in the data collection. Their recruitment was performed by the authors in their social and professional circles through word-of-mouth promotion. The data collection happened in several sessions, therefore some volunteers played both roles (\ie shopping assistant and client). The choice of product categories for the participants playing the shopping assistant is based on their self-assessment of their knowledge about these categories. After the data collection, participants were asked to complete an anonymous demographic survey to determine their general characteristics. Four dimensions are considered in the form:
\begin{enumerate*}
	\item gender,
	\item age,
	\item education,
	\item and the geographical origin of the participant's mother tongue.
\end{enumerate*} 
Table~\ref{tab:census} presents an overview by dimension of the 17 answers collected.
Seven females and ten males participated, with the majority being between 25 and 35 years old. The answers show that the linguistic background of the volunteers is diverse: they are from four different continents, although Europe and Asia are predominant. We hypothesize that this diversity can be reflected in the conversations, in their way of using English.
\begin{table}
    \vspace*{-\baselineskip}
	\centering
	\caption{Overview of census responses.}
    \captionshrink
	\label{tab:census}
	\begin{tabular}{l|l|c}
		\Xhline{1.2pt}
		Dimension & \multicolumn{2}{c}{Responses} \\ \Xhline{1.2pt}
		\multirow{2}{3cm}{Gender} & Female & 41.2\% \\
		& Male & 58.8\% \\ \Xhline{1.2pt}
		\multirow{2}{3cm}{Age} & 25-35 & 76.5\% \\
		& Over 35 & 23.5\% \\ \Xhline{1.2pt}
		\multirow{3}{3cm}{Education} & MSc & 47.1\% \\
		& PhD & 47.1\% \\
		& Other & 5.9\% \\ \Xhline{1.2pt}
		\multirow{4}{3cm}{Geographical continent of mother tongue} & Africa & 5.9\% \\
		& Asia & 41.2\% \\
		& Europe & 47.1\% \\ 
		& North America & 5.9\% \\ \Xhline{1.2pt}
	\end{tabular}
    \vspace*{-\baselineskip}
\end{table}

\subsection{Conversation Annotation}
\label{sec:collection:annotation}

In order to gain insights into the evolution of conversational goals along with the structure of the conversation, we annotate all utterances in terms of intents and goals.
For intent classification, we develop a schema based on previous work~\citep{Papenmeier:2022:CHIIR,Bunt:2017:chapter,Azzopardi:2018:CAIR}. For conversational goals, we use the three goals (QA, search, and recommendation), plus a fourth category for meta-communication. The schemata for intent and goal annotations are presented in Table~\ref{tab:annotation}.  
The advantage of these schemata are that they are generic and domain independent.

\begin{table*}
	\centering
	\caption{Schemata for intent (top) and goal annotation (bottom).}
    \captionshrink
	\label{tab:annotation}
	\begin{tabularx}{\linewidth}{ l | X }
		\Xhline{1.2pt}
		\textbf{Intent} & \textbf{Description} \\ \Xhline{1.2pt}
		Greetings & Indicates the beginning or end of the conversation \\ \hline
		Interaction structuring & Utterances that make the conversation structured and natural (e.g., thanking, stalling) \\ \hline
		Disclose & The client discloses information about what they are looking for \\ \hline
		Clarification question & The agent asks a question to make sure it understands correctly a previous statement \\ \hline
		Other question & Asks a question that is not a clarification question (e.g., factoid, follow-up questions) \\ \hline
		Elicit preferences & The agent asks a question to find the client's preferences (e.g., the color of an item, the budget) \\ \hline
		Recommend & The agent recommends one or several items to the client \\ \hline
		Answer & A participant gives an answer to the other participant's information request \\ \hline
		Explain &  Provides an explanation to a previous statement (e.g., justifies suggestion or rejection of an item) \\ \hline
		Positive feedback & Expresses positive feedback (e.g., confirmation, accept a recommendation) \\ \hline
		Negative feedback & Expresses negative feedback (e.g., disagreement, rejection of a recommendation) \\ \hline
		Other & Does not fit other labels \\ \Xhline{1.2pt}
		\textbf{Conversational goal} & \textbf{Description} \\ \Xhline{1.2pt}
		Search & The client wants to find more information on a product or specific topic. The agent answers the client’s request for information. This can take form of casual (why/how), unanswerable, or complex questions that require multiple interactions (e.g., follow-up, sub-questions) and their answers.  \\ \hline
		Recommendation & The agent elicits the client's preferences. The agent makes a recommendation based on the client's need and preferences. The client discloses what they are looking for or their preferences intentionally or as answer to the agent's questions. \\ \hline
		Question answering (QA) & A participant asks a factoid (what/when/who/where), confirmation (yes/no), or listing question about a product or specific topic. The other participant replies with a fact-based and short answer. \\ \hline
		Meta-communication & Makes the conversation fluid and natural but is not necessary to complete the goal of the conversation (\ie chit-chat). \\ \Xhline{1.2pt}
	\end{tabularx}
\end{table*}

\subsubsection{Intents}
The intent schema is based on a selection of communication functions from the international standard ISO 24617-2; these are domain independent and can help understand the participant's communicative behavior~\cite{Bunt:2017:chapter}. Similarly to~\citet{Papenmeier:2022:CHIIR}, we use only a limited number of generic intents from ISO 24617-2, as we do not know beforehand which intents will be present in the conversations. However, unlike them, we have several intents for inform, answer, and explain, as well as for the different types of questions such as clarification and preference elicitation~\citep{Azzopardi:2018:CAIR}.
In total, we select 12 diverse intents to characterize conversational patterns in the multi-goal conversations collected.
Indeed, some of these intents relate to the revealment  of information by the client or the shopping assistant (\eg inform, explain), while others such as positive and negative feedback represent the participants' sentiments.

To validate the intent schema, we compute the inter-annotator agreement between two experts annotators who perform intent annotation on three conversations. 
As a multi-annotator agreement measure of a multi-label task, we compute the Fleiss' kappa metric $\kappa$~\citep{Fleiss:1971:PsycholBull} per intent that we average to get a global inter-annotator agreement, as in~\citep{Budzianowski:2018:EMNLP}.
The agreement between the expert annotators is considered substantial ($\kappa$=0.633)~\citep{Landis:1977:biometrics}, which provides validation.

Intent annotation is carried out by crowd workers; for each conversation, five workers annotate every utterance and the intents selected by at least two annotators are kept. On the annotation user interface, the worker is shown an example along with the intent description table. Then, the conversation to annotate is displayed in a tabular form: the first column lists the utterances, the second shows the possible intents as checkboxes, and the last column has a text field where the annotator can optionally justify their choice.\footnote{A screenshot of the annotation user interface is included in our GitHub repository.} 
Table~\ref{tab:fleiss} shows that the inter-annotator agreement is lower between crowd workers ($\kappa$=0.187) than between experts. This can be explained by several factors, including the complexity of linguistic annotation tasks, and the number of annotators and labels~\citep{Artstein:2017:chapter}.

\begin{table}
	\centering
	\caption{Inter-annotator agreement and proportion of utterances per intent. Utterances can have multiple intent labels.}
    \captionshrink
	\label{tab:fleiss}
	\begin{tabular}{l|c|c}
		\Xhline{1.2pt}
		\textbf{Intent} & \textbf{Fleiss Kappa} & \textbf{\% utterances} \\ 
        \Xhline{1.2pt}
		Greetings & 0.434 & 14.3 \\ \hline
		Interaction structuring & 0.118 & 22.4 \\ \hline
		Disclose & 0.132 & 12.8 \\ \hline
		Clarification question & 0.159 & 23.4 \\ \hline
		Other question & 0.233 & 20.9 \\ \hline
		Elicit preferences & 0.085 & 11.2 \\ \hline
		Recommend & 0.176 & 14.3 \\ \hline
		Answer & 0.220 & 33.7 \\ \hline
		Explain & 0.153 & 22.7 \\ \hline
		Positive feedback & 0.155 & 16.5 \\ \hline
		Negative Feedback & 0.259 & 2.7 \\ \hline
		Other & 0.111 & 3.3 \\ \hline
		\Xhline{1.2pt}
		Weighted average & 0.187 & \\ \Xhline{1.2pt}
	\end{tabular}
\end{table}

\begin{figure*}
	\begin{subfigure}{0.48\textwidth}
        \vspace*{-\baselineskip}
		\includegraphics[width=\textwidth]{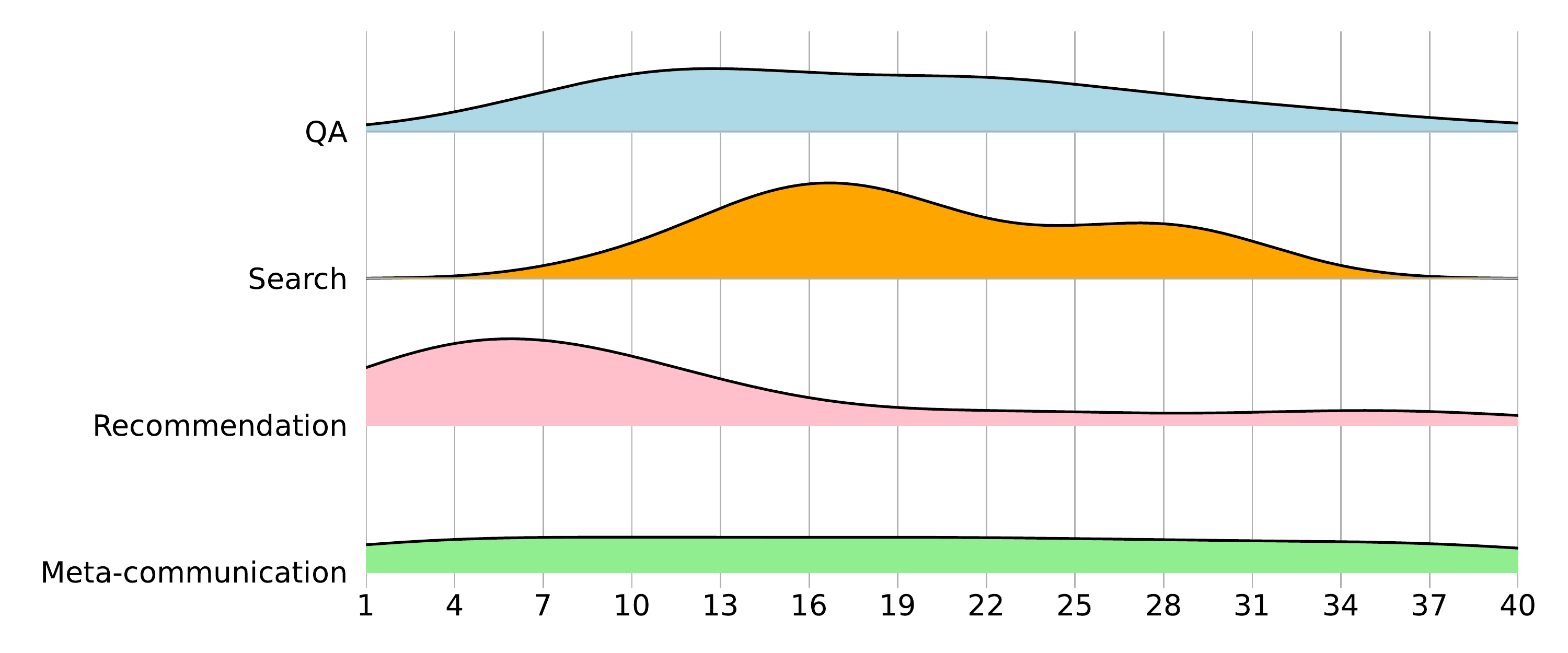}
        \vspace*{-\baselineskip}
		\caption{\textsc{Cell Phones and Accessories}, simple scenarios.}
		\label{fig:goals:simple}
	\end{subfigure}
	\begin{subfigure}{0.48\textwidth}
        \vspace*{-\baselineskip}
		\includegraphics[width=\textwidth]{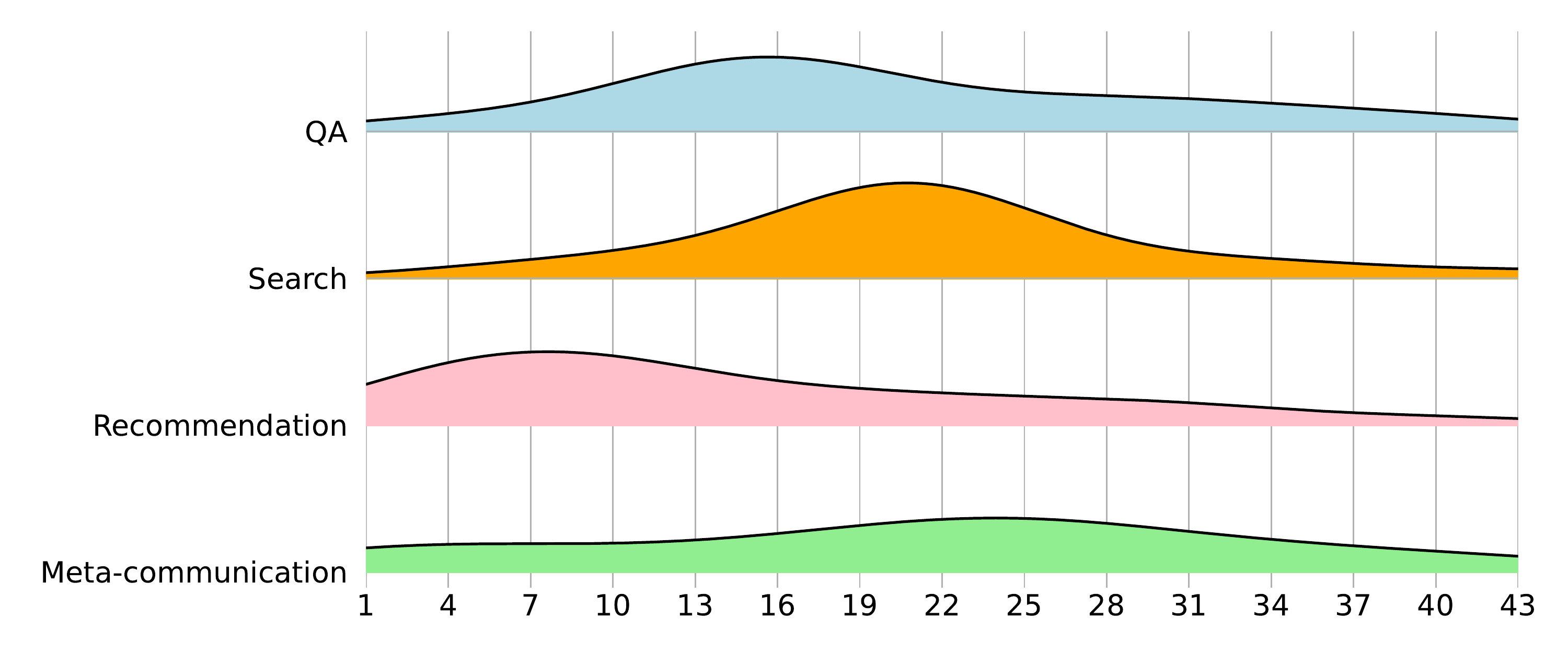}
        \vspace*{-\baselineskip}
		\caption{\textsc{Cell Phones and Accessories}, complex scenarios.}
		\label{fig:goals:complext}
	\end{subfigure}
	\begin{subfigure}{0.48\textwidth}
		\includegraphics[width=\textwidth]{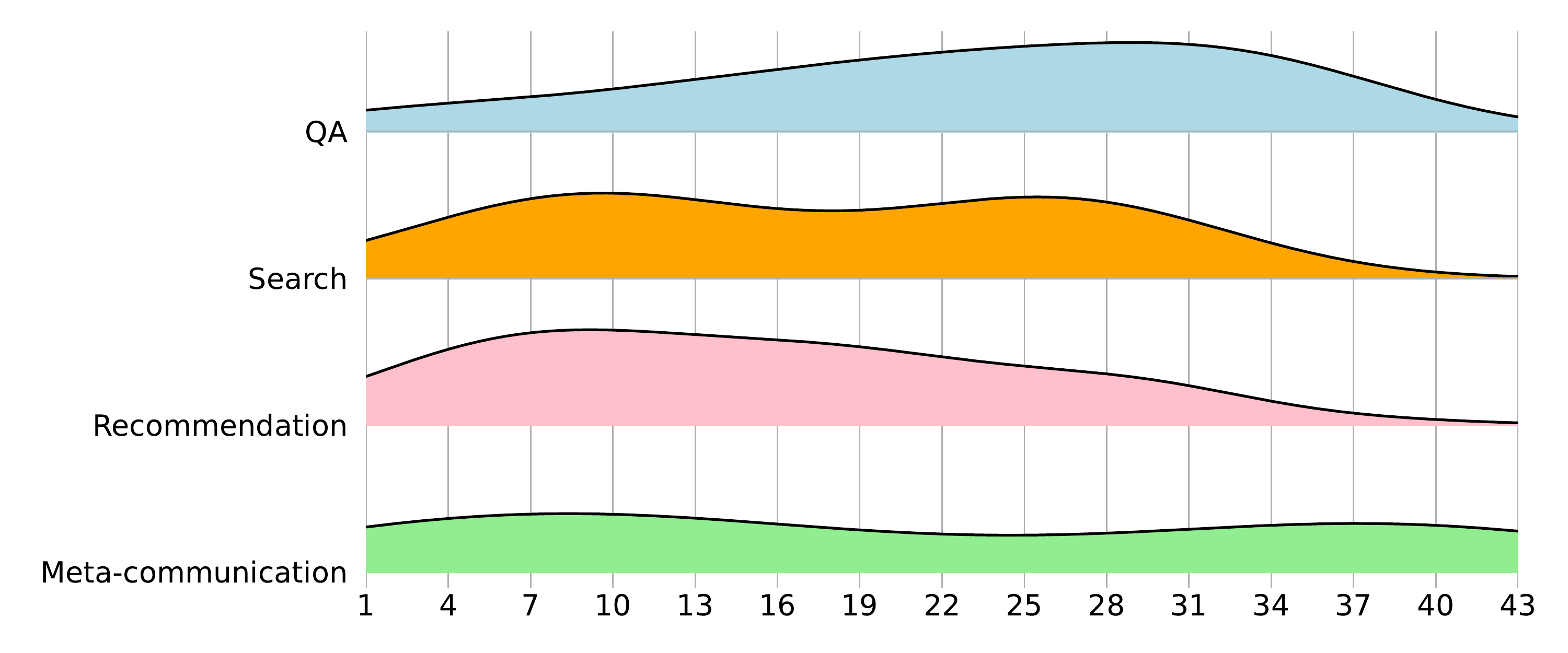}
        \vspace*{-\baselineskip}
		\caption{\textsc{Office Products}, simple scenarios.}
		\label{fig:goals:simple}
	\end{subfigure}
	\begin{subfigure}{0.48\textwidth}
		\includegraphics[width=\textwidth]{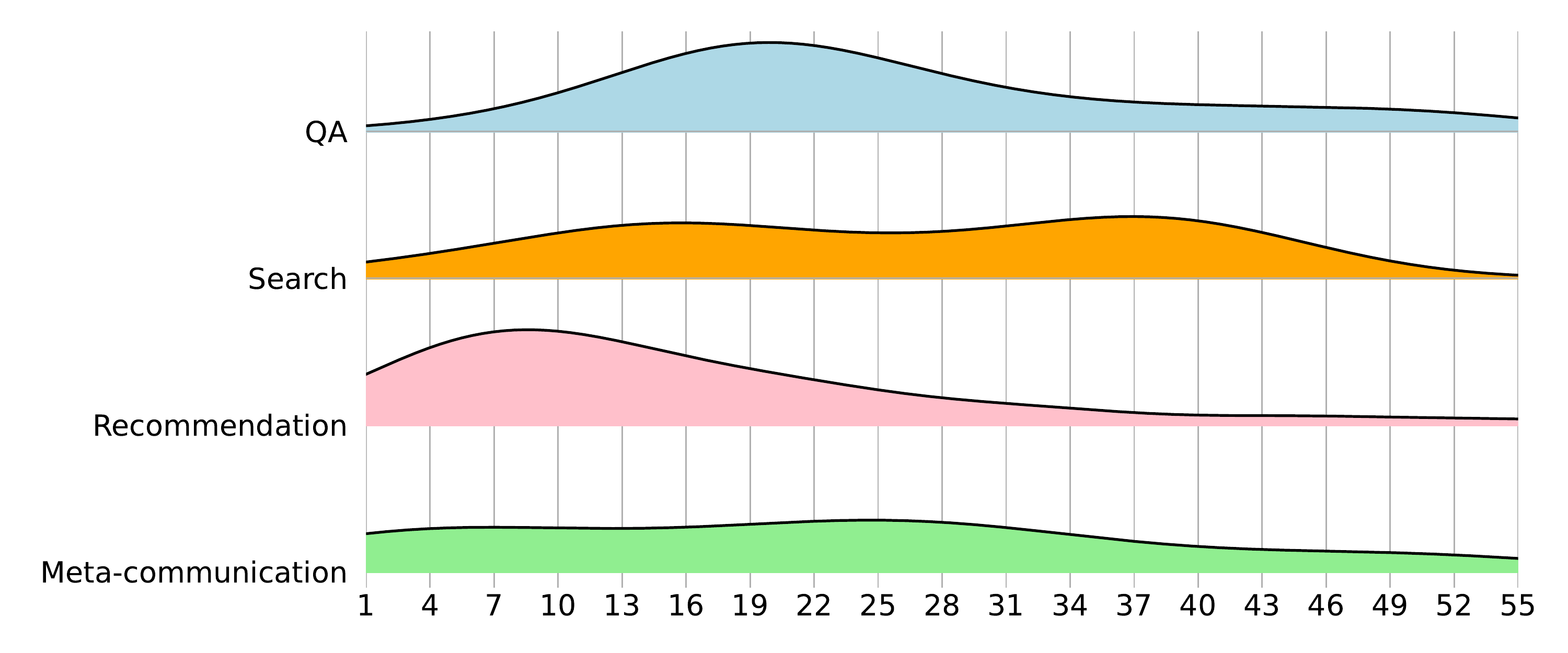}
        \vspace*{-\baselineskip}
		\caption{\textsc{Office Products}, complex scenarios.}
		\label{fig:goals:complext}
	\end{subfigure}
	\caption{Goal evolution for conversations about \textsc{Cell Phones and Accessories} (top) and \textsc{Office Products} (bottom).}
	\label{fig:goals}
\end{figure*}

\subsubsection{Goals}
Recall that we distinguish between three conversational goals: \emph{QA}, \emph{search}, and \emph{recommendation}. The delineation between the three conversational goals can be blurry,  especially for \emph{QA} and \emph{search}~\citep{Zamani:2022:arXiv}.
In the e-commerce context, we can easily imagine a conversation mixing these goals. For example, when looking for a book, one might ask more or less complex questions to a shopping assistant about the author or literary movement of a suggested book. 
Furthermore, we add meta-communication to this schema. Indeed, some utterances might not contribute to the completion of one of the three previous goals but structure the conversation and make it fluid.
Our motivation for annotating conversations with these goals is to observe whether some patterns emerge.
For this annotation task, we provide a detailed description for each goal in Table~\ref{tab:annotation}.
The description of \emph{recommendation} is based on the definitions in~\citep{Jannach:2021:CSUR,Gao:2021:AIOpen}; we emphasize on the preference elicitation element and the suggestion of products.
The distinction between \emph{QA} and \emph{search} is based on the type of question a participant can ask during the conversation~\citep{Zamani:2022:arXiv,Zaib:2022:KAIS}.
In this work, we consider QA questions to be short and factual, hence we look for the following types of questions: 
\begin{enumerate*} 
	\item factoid, commonly starting with interrogative words like what, when, where, and who (\eg ``What is the price of this book?'');
	\item confirmation (\eg ``Do you have it in blue?''); and
	\item listing (\eg ``Can you give me the specifications of this phone?'').
\end{enumerate*}
For \emph{search}, we have:
\begin{enumerate*}
	\item casual questions usually starting by why or how (\eg ``How are these shoes environmentally friendly?'');
	\item unanswerable questions; and 
	\item complex questions that require multiple interactions and can involve follow-up and subsequent questions (\eg ``What is the biggest difference between a beginner and a professional racquet? [...] Is there any difference in the string, like tension or wire thickness?'').
\end{enumerate*}

Goal annotation is done for every conversation by the first author of this paper. A test sample with 25\% of the conversations is also annotated by two crowd workers in order to ensure the clarity of the characterization of the goals. 
The crowdsourcing task is similar to the one for intent annotation, that is, an example is shown before starting the task, then the conversation is presented. However, instead of choosing one or multiple intents from a list, the crowd worker has to pick a single conversational goal.
We compute the Fleiss' kappa metric $\kappa$~\citep{Fleiss:1971:PsycholBull} to assess the agreement between the annotators. Despite the complexity and subjectivity of this annotation task, the inter-annotator agreement is moderate ($\kappa$=0.415)~\citep{Landis:1977:biometrics}, which provides validation for our schema.

\subsection{Dataset Summary}
\label{sec:collection:sum}

\begin{table}
	\centering
	\caption{Breakdown of MG-ShopDial per category.}
    \captionshrink
	\label{tab:categories}
	\begin{tabular}{ @{~~}l | @{~~}r@{~~} | @{~~}r@{~~~}}
		\Xhline{1.2pt}
		\textbf{Category} & \textbf{\#conversations} & \textbf{\#utterances} \\ \Xhline{1.2pt}
		\textsc{Books} & 14 & 482 \\ \hline
		\textsc{Office Products} & 14 & 458 \\ \hline
		\textsc{Sports and Outdoors} & 17 & 697 \\ \hline
		\textsc{Cell Phones and Accessories} & 19 & 559 \\ \Xhline{1.2pt}
		Total & 64 & 2,196 \\ \Xhline{1.2pt}
	\end{tabular}
\end{table}

The data collection procedure described above resulted in the creation of the MG-ShopDial collection, which contains a total of 64 conversations, comprising 2,196 utterances.
The number of conversations within each product category are reported in Table~\ref{tab:categories}. In general, we observe that the number of  conversation per category is almost balanced.
The complexity of the task is reflected by the number of utterances in a conversation: 75\% of the conversations have at least 23 utterances. On average, conversations have 34.3$\pm$14.9 utterances that are each 7.5$\pm$6.2 words long.

We observe that some conversations did not reach the end, \ie the selection of one or several products to buy. We hypothesize that some clients were more difficult to satisfy than others, requiring more utterances to understand their needs and elicit their preferences. However, the conversation is kept if it contains multiple goals. 
Furthermore, we notice that some conversations contain typos, grammatical errors, and emojis---these characteristics emphasize the naturalness of MG-ShopDial.
The conversations do not include personal information; if names were mentioned, they have been anonymized.

%% file: sigir2023-woz-06.tex
\section{Analysis}
\label{sec:analysis}

\begin{figure*}
    \centering
        \vspace*{-\baselineskip}
	\includegraphics[scale=0.30]{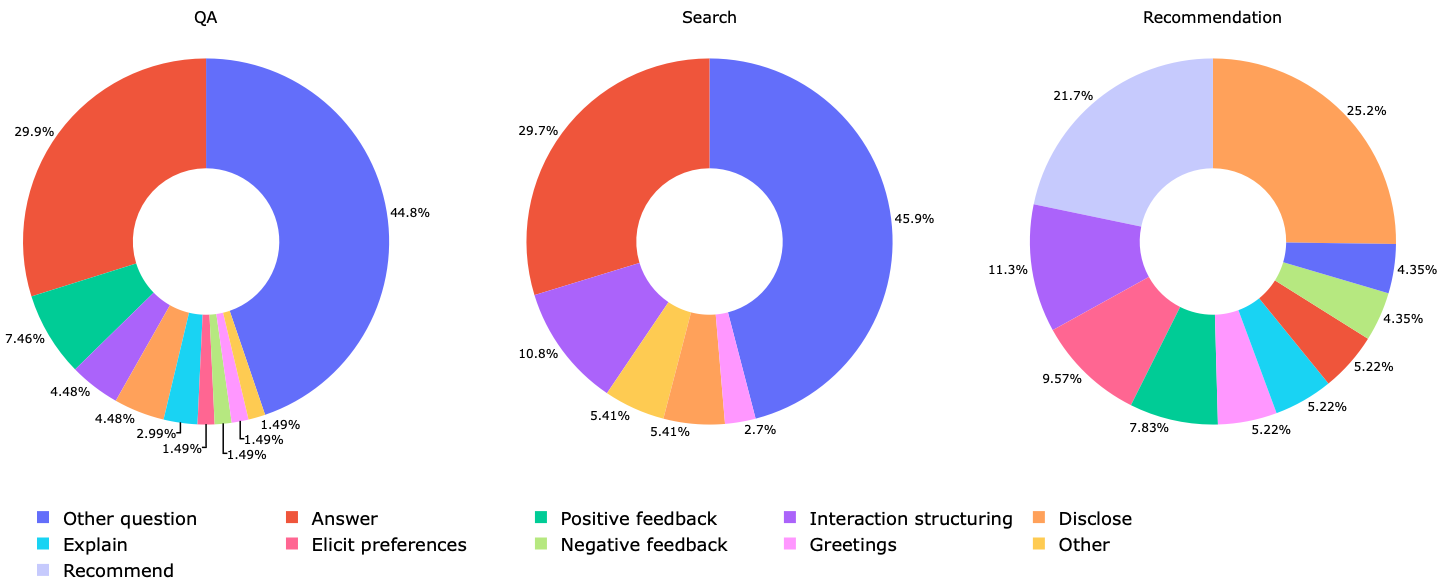}
        \vspace*{-0.75\baselineskip}
	\caption{Intent distribution per conversational goal on a MG-ShopDial sample.}
	\label{fig:correlation}
\end{figure*}

We analyze the MG-ShopDial dataset with a focus on the evolution of conversational goals (Section~\ref{sec:analysis:evo}) and the characterization of goals in terms of intents (Section~\ref{sec:analysis:cor}).

\subsection{Evolution of Conversational Goals}
\label{sec:analysis:evo}

To identify conversational patterns, we study the evolution of conversational goals per category and scenario complexity. 
We make the following main observations across the different categories:
\begin{itemize}[leftmargin=0.5cm]
    \item The difference in conversation length between the simple and complex scenarios is small, even though one could reasonably expect that finding multiple items would require more turns.
    \item The goal distributions are similar, despite different scenario complexities.
    \item A common trend for all conversation is to start with \emph{recommendation} followed by either \emph{search} or \emph{QA}, while having some \emph{meta-communication} all along. More specifically, the first part of the conversation focuses on uncovering the client's needs and preferences to recommend some products. The second part mostly consists of product-related information seeking. In some cases, we observe a second peak in \emph{recommendation} after the information seeking process, which might indicate that the client is not satisfied with the products and asks for other options.
\end{itemize}
Figure~\ref{fig:goals} illustrates these observations by displaying the distribution of goals over the course of the conversation for \textsc{Cell Phones and Accessories} and \textsc{Office Products}, under simple and complex scenarios.
The figure also shows the only exception to the main observations: simple scenarios for \textsc{Office Products}. There, the conversation length increases significantly for complex scenarios, and the goal distributions slightly deviate from the common trend. Indeed, the distributions are more balanced, yet we can still observe a decreasing probability for \emph{recommendation} and the opposite for \emph{QA} as the conversation progresses. Further, we note that \emph{search} is present almost uniformly for most of the duration of the conversation. This might indicate that participants found this category more conducive to searching.

The different observations are consistent with some of the findings on product search by \citet{Papenmeier:2022:CHIIR}. Specifically, the almost uniform distribution of \emph{meta-communication} illustrates the need of supporting interaction structuring (\eg stalling, thanking).  Further, the prevalence of \emph{recommendation} over the first 10-15 utterances supports the idea of strategically uncovering and narrowing the client's needs and preferences.

\subsection{Intent-based Characterization of Goals}
\label{sec:analysis:cor}

Figure~\ref{fig:correlation} shows the intent distribution per goal on a sample of conversations annotated with goals and intents by the main author (to ensure consistency and reduce potential noise from crowd annotations).
For \emph{QA} and \emph{search}, \emph{Answer} and \emph{Other question} represent around 75\% of the intents present, which is consistent with the idea of asking questions to get more information on products. Also, the \emph{Recommend} intent is not present at all for these goals.
As it can be expected for \emph{recommendation}, the intents \emph{Recommend} and \emph{Disclose} represent the majority of the annotations (46.9\%), followed by \emph{Interaction structuring} (11.3\%) and \emph{Preference elicitation} (9.6\%). 
These observations are in accord with the ambiguity around the different conversational goals, esp. with regards to \emph{QA} and \emph{search}. 
Indeed, \emph{recommendation} can be easily distinguished from \emph{QA} and \emph{search} based on the intent distributions, while the distinction between the latter two is far less obvious.
An idea for future work is to refine the \emph{Other question} and \emph{Answer} intents to see whether that would yield a better differentiation between \emph{QA} and \emph{search}.

%% file: sigir2023-woz-07.tex
\section{Conclusion}
\label{sec:concl}

In this work, we have introduced the MG-ShopDial dataset, along with the resources used to create it.
Specifically, we have proposed a coached human-human protocol that emphasizes on guiding participants with the help of checklists instead of giving them a rigid script, and have developed the Coached Conversation Collector tool to perform the data collection following this protocol. 
The collected data has been annotated on the utterance level with both intents and conversational goals.
Upon analyzing MG-ShopDial, we have observed a consistent conversational pattern that typically involved two or three distinct phases: initially, a recommendation is made, followed by information seeking, and in some instances, a secondary recommendation. 
We have also noticed that meta-communication is used throughout the conversation to keep it natural and to help transition between the different goals.
Finally, the characterization of conversational goals in terms of intents has shown a clear distinction between \emph{recommendation} and \emph{search}/\emph{QA}, but not so much between the latter two.

To the best of our knowledge, MG-ShopDial is the first dataset that mixes multiple conversational goals in a natural manner by situating participants in an e-commerce scenario. As such, it allows the development of conversational agents that support multiple goals. Nonetheless, the dataset is too small in size to train agents in an end-to-end manner. One solution would be to collect more data using our protocol and tool. However, creating a large collection with the same quality standards as ours is likely to be very time consuming and expensive. 
Another use of the dataset could be for few-shot learning with newer large language models, such as GPT-4~\citep{Openai:2023:arXiv}.
Alternatively, one could employ user simulation~\citep{Balog:2021:DESIRES}; the collection is large enough to learn the parameters of models that can effectively characterize different scenarios.

%% file: 00paper.bbl

\begin{thebibliography}{49}


\ifx \showCODEN    \undefined \def \showCODEN     #1{\unskip}     \fi
\ifx \showDOI      \undefined \def \showDOI       #1{#1}\fi
\ifx \showISBNx    \undefined \def \showISBNx     #1{\unskip}     \fi
\ifx \showISBNxiii \undefined \def \showISBNxiii  #1{\unskip}     \fi
\ifx \showISSN     \undefined \def \showISSN      #1{\unskip}     \fi
\ifx \showLCCN     \undefined \def \showLCCN      #1{\unskip}     \fi
\ifx \shownote     \undefined \def \shownote      #1{#1}          \fi
\ifx \showarticletitle \undefined \def \showarticletitle #1{#1}   \fi
\ifx \showURL      \undefined \def \showURL       {\relax}        \fi
\providecommand\bibfield[2]{#2}
\providecommand\bibinfo[2]{#2}
\providecommand\natexlab[1]{#1}
\providecommand\showeprint[2][]{arXiv:#2}

\bibitem[\protect\citeauthoryear{Allouch, Azaria, and Azoulay}{Allouch
  et~al\mbox{.}}{2021}]%
        {Allouch:2021:Sensors}
\bibfield{author}{\bibinfo{person}{Merav Allouch}, \bibinfo{person}{Amos
  Azaria}, {and} \bibinfo{person}{Rina Azoulay}.}
  \bibinfo{year}{2021}\natexlab{}.
\newblock \showarticletitle{Conversational Agents: Goals, Technologies, Vision
  and Challenges}.
\newblock \bibinfo{journal}{\emph{Sensors}} \bibinfo{volume}{21},
  \bibinfo{number}{24} (\bibinfo{year}{2021}).
\newblock


\bibitem[\protect\citeauthoryear{Artstein}{Artstein}{2017}]%
        {Artstein:2017:chapter}
\bibfield{author}{\bibinfo{person}{Ron Artstein}.}
  \bibinfo{year}{2017}\natexlab{}.
\newblock \bibinfo{booktitle}{\emph{Inter-annotator Agreement}}.
\newblock \bibinfo{publisher}{Springer Netherlands}, \bibinfo{pages}{297--313}.
\newblock


\bibitem[\protect\citeauthoryear{Azzopardi, Dubiel, Halvey, and
  Dalton}{Azzopardi et~al\mbox{.}}{2018}]%
        {Azzopardi:2018:CAIR}
\bibfield{author}{\bibinfo{person}{Leif Azzopardi}, \bibinfo{person}{Mateusz
  Dubiel}, \bibinfo{person}{Martin Halvey}, {and} \bibinfo{person}{Jeffery
  Dalton}.} \bibinfo{year}{2018}\natexlab{}.
\newblock \showarticletitle{Conceptualizing agent-human interactions during the
  conversational search process}. In \bibinfo{booktitle}{\emph{SIGIR 2nd
  International Workshop on Conversational Approaches to Information
  Retrieval}} \emph{(\bibinfo{series}{CAIR '18})}.
\newblock


\bibitem[\protect\citeauthoryear{Baheti, Ritter, and Small}{Baheti
  et~al\mbox{.}}{2020}]%
        {Baheti:2020:ACL}
\bibfield{author}{\bibinfo{person}{Ashutosh Baheti}, \bibinfo{person}{Alan
  Ritter}, {and} \bibinfo{person}{Kevin Small}.}
  \bibinfo{year}{2020}\natexlab{}.
\newblock \showarticletitle{Fluent Response Generation for Conversational
  Question Answering}. In \bibinfo{booktitle}{\emph{Proceedings of the 58th
  Annual Meeting of the Association for Computational Linguistics}}.
  \bibinfo{pages}{191--207}.
\newblock


\bibitem[\protect\citeauthoryear{Bajaj, Campos, Craswell, Deng, Gao, Liu,
  Majumder, McNamara, Mitra, Nguyen, Rosenberg, Song, Stoica, Tiwary, and
  Wang}{Bajaj et~al\mbox{.}}{2016}]%
        {Bajaj:2016:arXiv}
\bibfield{author}{\bibinfo{person}{Payal Bajaj}, \bibinfo{person}{Daniel
  Campos}, \bibinfo{person}{Nick Craswell}, \bibinfo{person}{Li Deng},
  \bibinfo{person}{Jianfeng Gao}, \bibinfo{person}{Xiaodong Liu},
  \bibinfo{person}{Rangan Majumder}, \bibinfo{person}{Andrew McNamara},
  \bibinfo{person}{Bhaskar Mitra}, \bibinfo{person}{Tri Nguyen},
  \bibinfo{person}{Mir Rosenberg}, \bibinfo{person}{Xia Song},
  \bibinfo{person}{Alina Stoica}, \bibinfo{person}{Saurabh Tiwary}, {and}
  \bibinfo{person}{Tong Wang}.} \bibinfo{year}{2016}\natexlab{}.
\newblock \bibinfo{title}{MS MARCO: A Human Generated MAchine Reading
  COmprehension Dataset}.
\newblock
\newblock
\showeprint[arxiv]{1611.09268}~[cs.CL]


\bibitem[\protect\citeauthoryear{Balog}{Balog}{2021}]%
        {Balog:2021:DESIRES}
\bibfield{author}{\bibinfo{person}{Krisztian Balog}.}
  \bibinfo{year}{2021}\natexlab{}.
\newblock \showarticletitle{Conversational {AI} from an Information Retrieval
  Perspective: {R}emaining Challenges and a Case for User Simulation}. In
  \bibinfo{booktitle}{\emph{Proceedings of the 2nd International Conference on
  Design of Experimental Search \& Information REtrieval Systems}}
  \emph{(\bibinfo{series}{DESIRES '21})}. \bibinfo{pages}{80--90}.
\newblock


\bibitem[\protect\citeauthoryear{Bennett and Rudnicky}{Bennett and
  Rudnicky}{2002}]%
        {Bennett:2002:ICSLP}
\bibfield{author}{\bibinfo{person}{Christina Bennett} {and}
  \bibinfo{person}{Alexander~I. Rudnicky}.} \bibinfo{year}{2002}\natexlab{}.
\newblock \showarticletitle{The Carnegie Mellon Communicator Corpus}. In
  \bibinfo{booktitle}{\emph{Proc. 7th International Conference on Spoken
  Language Processing}} \emph{(\bibinfo{series}{ICSLP '02})}.
  \bibinfo{pages}{341--344}.
\newblock


\bibitem[\protect\citeauthoryear{Budzianowski, Wen, Tseng, Casanueva, Ultes,
  Ramadan, and Ga{\v{s}}i{\'c}}{Budzianowski et~al\mbox{.}}{2018}]%
        {Budzianowski:2018:EMNLP}
\bibfield{author}{\bibinfo{person}{Pawe{\l} Budzianowski},
  \bibinfo{person}{Tsung-Hsien Wen}, \bibinfo{person}{Bo-Hsiang Tseng},
  \bibinfo{person}{I{\~n}igo Casanueva}, \bibinfo{person}{Stefan Ultes},
  \bibinfo{person}{Osman Ramadan}, {and} \bibinfo{person}{Milica
  Ga{\v{s}}i{\'c}}.} \bibinfo{year}{2018}\natexlab{}.
\newblock \showarticletitle{MultiWOZ - A Large-Scale Multi-Domain Wizard-of-Oz
  Dataset for Task-Oriented Dialogue Modelling}. In
  \bibinfo{booktitle}{\emph{Proceedings of the 2018 Conference on Empirical
  Methods in Natural Language Processing}} \emph{(\bibinfo{series}{EMNLP
  '18})}. \bibinfo{pages}{5016--5026}.
\newblock


\bibitem[\protect\citeauthoryear{Bunt, Petukhova, Traum, and
  Alexandersson}{Bunt et~al\mbox{.}}{2017}]%
        {Bunt:2017:chapter}
\bibfield{author}{\bibinfo{person}{Harry Bunt}, \bibinfo{person}{Volha
  Petukhova}, \bibinfo{person}{David Traum}, {and} \bibinfo{person}{Jan
  Alexandersson}.} \bibinfo{year}{2017}\natexlab{}.
\newblock \showarticletitle{Dialogue act annotation with the ISO 24617-2
  standard}.
\newblock In \bibinfo{booktitle}{\emph{Multimodal interaction with W3C
  standards}}. \bibinfo{pages}{109--135}.
\newblock


\bibitem[\protect\citeauthoryear{Cai and Chen}{Cai and Chen}{2020}]%
        {Cai:2020:UMAP}
\bibfield{author}{\bibinfo{person}{Wanling Cai} {and} \bibinfo{person}{Li
  Chen}.} \bibinfo{year}{2020}\natexlab{}.
\newblock \showarticletitle{Predicting User Intents and Satisfaction with
  Dialogue-Based Conversational Recommendations}. In
  \bibinfo{booktitle}{\emph{Proceedings of the 28th ACM Conference on User
  Modeling, Adaptation and Personalization}} \emph{(\bibinfo{series}{UMAP
  '20})}. \bibinfo{pages}{33--42}.
\newblock


\bibitem[\protect\citeauthoryear{Campos, Otegi, Soroa, Deriu, Cieliebak, and
  Agirre}{Campos et~al\mbox{.}}{2020}]%
        {Campos:2020:ACL}
\bibfield{author}{\bibinfo{person}{Jon~Ander Campos}, \bibinfo{person}{Arantxa
  Otegi}, \bibinfo{person}{Aitor Soroa}, \bibinfo{person}{Jan Deriu},
  \bibinfo{person}{Mark Cieliebak}, {and} \bibinfo{person}{Eneko Agirre}.}
  \bibinfo{year}{2020}\natexlab{}.
\newblock \showarticletitle{DoQA - Accessing Domain-Specific FAQs via
  Conversational QA}. In \bibinfo{booktitle}{\emph{Proceedings of the 58th
  Annual Meeting of the Association for Computational Linguistics}}
  \emph{(\bibinfo{series}{ACL '20})}. \bibinfo{pages}{7302--7314}.
\newblock


\bibitem[\protect\citeauthoryear{Choi, He, Iyyer, Yatskar, Yih, Choi, Liang,
  and Zettlemoyer}{Choi et~al\mbox{.}}{2018}]%
        {Choi:2018:EMNLP}
\bibfield{author}{\bibinfo{person}{Eunsol Choi}, \bibinfo{person}{He He},
  \bibinfo{person}{Mohit Iyyer}, \bibinfo{person}{Mark Yatskar},
  \bibinfo{person}{Wen-tau Yih}, \bibinfo{person}{Yejin Choi},
  \bibinfo{person}{Percy Liang}, {and} \bibinfo{person}{Luke Zettlemoyer}.}
  \bibinfo{year}{2018}\natexlab{}.
\newblock \showarticletitle{QuAC: Question Answering in Context}. In
  \bibinfo{booktitle}{\emph{Proceedings of the 2018 Conference on Empirical
  Methods in Natural Language Processing}} \emph{(\bibinfo{series}{EMNLP
  '18})}. \bibinfo{pages}{2174--2184}.
\newblock


\bibitem[\protect\citeauthoryear{Culpepper, Diaz, and Smucker}{Culpepper
  et~al\mbox{.}}{2018}]%
        {Culpepper:2018:SIGIRForum}
\bibfield{author}{\bibinfo{person}{J.~Shane Culpepper},
  \bibinfo{person}{Fernando Diaz}, {and} \bibinfo{person}{Mark~D. Smucker}.}
  \bibinfo{year}{2018}\natexlab{}.
\newblock \showarticletitle{Research Frontiers in Information Retrieval: Report
  from the Third Strategic Workshop on Information Retrieval in Lorne (SWIRL
  2018)}.
\newblock \bibinfo{journal}{\emph{SIGIR Forum}} \bibinfo{volume}{52},
  \bibinfo{number}{1} (\bibinfo{year}{2018}), \bibinfo{pages}{34––90}.
\newblock


\bibitem[\protect\citeauthoryear{Dalton, Xiong, Kumar, and Callan}{Dalton
  et~al\mbox{.}}{2020}]%
        {Dalton:2020:SIGIR}
\bibfield{author}{\bibinfo{person}{Jeff Dalton}, \bibinfo{person}{Chenyan
  Xiong}, \bibinfo{person}{Vaibhav Kumar}, {and} \bibinfo{person}{Jamie
  Callan}.} \bibinfo{year}{2020}\natexlab{}.
\newblock \showarticletitle{CAsT-19: A Dataset for Conversational Information
  Seeking}. In \bibinfo{booktitle}{\emph{Proceedings of the 43rd International
  ACM SIGIR Conference on Research and Development in Information Retrieval}}
  \emph{(\bibinfo{series}{SIGIR '20})}. \bibinfo{pages}{1985--1988}.
\newblock


\bibitem[\protect\citeauthoryear{Dodge, Gane, Zhang, Bordes, Chopra, Miller,
  Szlam, and Weston}{Dodge et~al\mbox{.}}{2016}]%
        {Dodge:2016:ICLR}
\bibfield{author}{\bibinfo{person}{Jesse Dodge}, \bibinfo{person}{Andreea
  Gane}, \bibinfo{person}{Xiang Zhang}, \bibinfo{person}{Antoine Bordes},
  \bibinfo{person}{Sumit Chopra}, \bibinfo{person}{{Alexander H.} Miller},
  \bibinfo{person}{Arthur Szlam}, {and} \bibinfo{person}{Jason Weston}.}
  \bibinfo{year}{2016}\natexlab{}.
\newblock \showarticletitle{Evaluating prerequisite qualities for learning
  end-to-end dialog systems}. In \bibinfo{booktitle}{\emph{4th International
  Conference on Learning Representations}} \emph{(\bibinfo{series}{ICLR '16})}.
\newblock


\bibitem[\protect\citeauthoryear{Fleiss}{Fleiss}{1971}]%
        {Fleiss:1971:PsycholBull}
\bibfield{author}{\bibinfo{person}{Joseph~L Fleiss}.}
  \bibinfo{year}{1971}\natexlab{}.
\newblock \showarticletitle{Measuring nominal scale agreement among many
  raters.}
\newblock \bibinfo{journal}{\emph{Psychological bulletin}}
  \bibinfo{volume}{76}, \bibinfo{number}{5} (\bibinfo{year}{1971}).
\newblock


\bibitem[\protect\citeauthoryear{Gao, Lei, He, {de Rijke}, and Chua}{Gao
  et~al\mbox{.}}{2021}]%
        {Gao:2021:AIOpen}
\bibfield{author}{\bibinfo{person}{Chongming Gao}, \bibinfo{person}{Wenqiang
  Lei}, \bibinfo{person}{Xiangnan He}, \bibinfo{person}{Maarten {de Rijke}},
  {and} \bibinfo{person}{Tat-Seng Chua}.} \bibinfo{year}{2021}\natexlab{}.
\newblock \showarticletitle{Advances and challenges in conversational
  recommender systems: A survey}.
\newblock \bibinfo{journal}{\emph{AI Open}}  \bibinfo{volume}{2}
  (\bibinfo{year}{2021}), \bibinfo{pages}{100--126}.
\newblock


\bibitem[\protect\citeauthoryear{Gao, Xiong, Bennett, and Craswell}{Gao
  et~al\mbox{.}}{2022}]%
        {Gao:2022:arXiv}
\bibfield{author}{\bibinfo{person}{Jianfeng Gao}, \bibinfo{person}{Chenyan
  Xiong}, \bibinfo{person}{Paul Bennett}, {and} \bibinfo{person}{Nick
  Craswell}.} \bibinfo{year}{2022}\natexlab{}.
\newblock \bibinfo{title}{Neural Approaches to Conversational Information
  Retrieval}.
\newblock
\newblock
\showeprint[arxiv]{2201.05176}~[cs.IR]


\bibitem[\protect\citeauthoryear{Hayati, Kang, Zhu, Shi, and Yu}{Hayati
  et~al\mbox{.}}{2020}]%
        {Hayati:2020:EMNLP}
\bibfield{author}{\bibinfo{person}{Shirley~Anugrah Hayati},
  \bibinfo{person}{Dongyeop Kang}, \bibinfo{person}{Qingxiaoyang Zhu},
  \bibinfo{person}{Weiyan Shi}, {and} \bibinfo{person}{Zhou Yu}.}
  \bibinfo{year}{2020}\natexlab{}.
\newblock \showarticletitle{INSPIRED: Toward Sociable Recommendation Dialog
  Systems}. In \bibinfo{booktitle}{\emph{Proceedings of the 2020 Conference on
  Empirical Methods in Natural Language Processing}}
  \emph{(\bibinfo{series}{EMNLP '20})}. \bibinfo{pages}{8142--8152}.
\newblock


\bibitem[\protect\citeauthoryear{He, Balakrishnan, Eric, and Liang}{He
  et~al\mbox{.}}{2017}]%
        {He:2017:ACL}
\bibfield{author}{\bibinfo{person}{He He}, \bibinfo{person}{Anusha
  Balakrishnan}, \bibinfo{person}{Mihail Eric}, {and} \bibinfo{person}{Percy
  Liang}.} \bibinfo{year}{2017}\natexlab{}.
\newblock \showarticletitle{Learning Symmetric Collaborative Dialogue Agents
  with Dynamic Knowledge Graph Embeddings}. In
  \bibinfo{booktitle}{\emph{Proceedings of the 55th Annual Meeting of the
  Association for Computational Linguistics (Volume 1: Long Papers)}}
  \emph{(\bibinfo{series}{ACL '17})}. \bibinfo{pages}{1766--1776}.
\newblock


\bibitem[\protect\citeauthoryear{He, Chen, Balakrishnan, and Liang}{He
  et~al\mbox{.}}{2018}]%
        {He:2018:EMNLP}
\bibfield{author}{\bibinfo{person}{He He}, \bibinfo{person}{Derek Chen},
  \bibinfo{person}{Anusha Balakrishnan}, {and} \bibinfo{person}{Percy Liang}.}
  \bibinfo{year}{2018}\natexlab{}.
\newblock \showarticletitle{Decoupling Strategy and Generation in Negotiation
  Dialogues}. In \bibinfo{booktitle}{\emph{Proceedings of the 2018 Conference
  on Empirical Methods in Natural Language Processing}}
  \emph{(\bibinfo{series}{EMNLP '18})}. \bibinfo{pages}{2333--2343}.
\newblock


\bibitem[\protect\citeauthoryear{Jannach, Manzoor, Cai, and Chen}{Jannach
  et~al\mbox{.}}{2021}]%
        {Jannach:2021:CSUR}
\bibfield{author}{\bibinfo{person}{Dietmar Jannach}, \bibinfo{person}{Ahtsham
  Manzoor}, \bibinfo{person}{Wanling Cai}, {and} \bibinfo{person}{Li Chen}.}
  \bibinfo{year}{2021}\natexlab{}.
\newblock \showarticletitle{A Survey on Conversational Recommender Systems}.
\newblock \bibinfo{journal}{\emph{ACM Comput. Surv.}} \bibinfo{volume}{54},
  \bibinfo{number}{5} (\bibinfo{year}{2021}), \bibinfo{pages}{1--36}.
\newblock


\bibitem[\protect\citeauthoryear{Joko, Hasibi, Balog, and de~Vries}{Joko
  et~al\mbox{.}}{2021}]%
        {Joko:2021:SIGIR}
\bibfield{author}{\bibinfo{person}{Hideaki Joko}, \bibinfo{person}{Faegheh
  Hasibi}, \bibinfo{person}{Krisztian Balog}, {and} \bibinfo{person}{Arjen~P.
  de Vries}.} \bibinfo{year}{2021}\natexlab{}.
\newblock \showarticletitle{Conversational Entity Linking: Problem Definition
  and Datasets}. In \bibinfo{booktitle}{\emph{Proceedings of the 44th
  International ACM SIGIR Conference on Research and Development in Information
  Retrieval}} \emph{(\bibinfo{series}{SIGIR '21})}.
  \bibinfo{pages}{2390--2397}.
\newblock


\bibitem[\protect\citeauthoryear{Kostric, Balog, and Radlinski}{Kostric
  et~al\mbox{.}}{2021}]%
        {Kostric:2021:RecSys}
\bibfield{author}{\bibinfo{person}{Ivica Kostric}, \bibinfo{person}{Krisztian
  Balog}, {and} \bibinfo{person}{Filip Radlinski}.}
  \bibinfo{year}{2021}\natexlab{}.
\newblock \showarticletitle{Soliciting User Preferences in Conversational
  Recommender Systems via Usage-Related Questions}. In
  \bibinfo{booktitle}{\emph{Fifteenth ACM Conference on Recommender Systems}}
  \emph{(\bibinfo{series}{RecSys '21})}. \bibinfo{pages}{724--729}.
\newblock
\urldef\tempurl%
\url{https://doi.org/10.1145/3460231.3478861}
\showDOI{\tempurl}


\bibitem[\protect\citeauthoryear{Landis and Koch}{Landis and Koch}{1977}]%
        {Landis:1977:biometrics}
\bibfield{author}{\bibinfo{person}{J~Richard Landis} {and}
  \bibinfo{person}{Gary~G Koch}.} \bibinfo{year}{1977}\natexlab{}.
\newblock \showarticletitle{The measurement of observer agreement for
  categorical data}.
\newblock \bibinfo{journal}{\emph{Biometrics}} \bibinfo{volume}{33},
  \bibinfo{number}{1} (\bibinfo{year}{1977}), \bibinfo{pages}{159--174}.
\newblock


\bibitem[\protect\citeauthoryear{Li, Kahou, Schulz, Michalski, Charlin, and
  Pal}{Li et~al\mbox{.}}{2018}]%
        {Li:2018:NIPS}
\bibfield{author}{\bibinfo{person}{Raymond Li}, \bibinfo{person}{Samira Kahou},
  \bibinfo{person}{Hannes Schulz}, \bibinfo{person}{Vincent Michalski},
  \bibinfo{person}{Laurent Charlin}, {and} \bibinfo{person}{Chris Pal}.}
  \bibinfo{year}{2018}\natexlab{}.
\newblock \showarticletitle{Towards Deep Conversational Recommendations}. In
  \bibinfo{booktitle}{\emph{Proceedings of the 32nd International Conference on
  Neural Information Processing Systems}} \emph{(\bibinfo{series}{NIPS '18})}.
  \bibinfo{pages}{9748--9758}.
\newblock


\bibitem[\protect\citeauthoryear{Lin, Yang, Nogueira, Tsai, Wang, and Lin}{Lin
  et~al\mbox{.}}{2021}]%
        {Lin:2021:TOIS}
\bibfield{author}{\bibinfo{person}{Sheng-Chieh Lin},
  \bibinfo{person}{Jheng-Hong Yang}, \bibinfo{person}{Rodrigo Nogueira},
  \bibinfo{person}{Ming-Feng Tsai}, \bibinfo{person}{Chuan-Ju Wang}, {and}
  \bibinfo{person}{Jimmy Lin}.} \bibinfo{year}{2021}\natexlab{}.
\newblock \showarticletitle{Multi-stage conversational passage retrieval: An
  approach to fusing term importance estimation and neural query rewriting}.
\newblock \bibinfo{journal}{\emph{ACM Transactions on Information Systems}}
  \bibinfo{volume}{39}, \bibinfo{number}{4} (\bibinfo{year}{2021}),
  \bibinfo{pages}{1--29}.
\newblock


\bibitem[\protect\citeauthoryear{Liu, Wang, Niu, Wu, and Che}{Liu
  et~al\mbox{.}}{2021}]%
        {Liu:2021:EMNLP}
\bibfield{author}{\bibinfo{person}{Zeming Liu}, \bibinfo{person}{Haifeng Wang},
  \bibinfo{person}{Zheng-Yu Niu}, \bibinfo{person}{Hua Wu}, {and}
  \bibinfo{person}{Wanxiang Che}.} \bibinfo{year}{2021}\natexlab{}.
\newblock \showarticletitle{DuRecDial 2.0: A Bilingual Parallel Corpus for
  Conversational Recommendation}. In \bibinfo{booktitle}{\emph{Proceedings of
  the 2021 Conference on Empirical Methods in Natural Language Processing}}
  \emph{(\bibinfo{series}{EMNLP '21})}. \bibinfo{pages}{4335--4347}.
\newblock


\bibitem[\protect\citeauthoryear{Liu, Wang, Niu, Wu, Che, and Liu}{Liu
  et~al\mbox{.}}{2020}]%
        {Liu:2020:ACL}
\bibfield{author}{\bibinfo{person}{Zeming Liu}, \bibinfo{person}{Haifeng Wang},
  \bibinfo{person}{Zheng-Yu Niu}, \bibinfo{person}{Hua Wu},
  \bibinfo{person}{Wanxiang Che}, {and} \bibinfo{person}{Ting Liu}.}
  \bibinfo{year}{2020}\natexlab{}.
\newblock \showarticletitle{Towards Conversational Recommendation over
  Multi-Type Dialogs}. In \bibinfo{booktitle}{\emph{Proceedings of the 58th
  Annual Meeting of the Association for Computational Linguistics}}
  \emph{(\bibinfo{series}{ACL '20})}. \bibinfo{pages}{1036--1049}.
\newblock


\bibitem[\protect\citeauthoryear{Miller, Feng, Fisch, Lu, Batra, Bordes,
  Parikh, and Weston}{Miller et~al\mbox{.}}{2017}]%
        {Miller:2017:arXiv}
\bibfield{author}{\bibinfo{person}{Alexander~H. Miller}, \bibinfo{person}{Will
  Feng}, \bibinfo{person}{Adam Fisch}, \bibinfo{person}{Jiasen Lu},
  \bibinfo{person}{Dhruv Batra}, \bibinfo{person}{Antoine Bordes},
  \bibinfo{person}{Devi Parikh}, {and} \bibinfo{person}{Jason Weston}.}
  \bibinfo{year}{2017}\natexlab{}.
\newblock \bibinfo{title}{ParlAI: A Dialog Research Software Platform}.
\newblock
\newblock
\showeprint[arxiv]{1705.06476}~[cs.CL]


\bibitem[\protect\citeauthoryear{Ni, Li, and McAuley}{Ni et~al\mbox{.}}{2019}]%
        {Ni:2019:EMNLP}
\bibfield{author}{\bibinfo{person}{Jianmo Ni}, \bibinfo{person}{Jiacheng Li},
  {and} \bibinfo{person}{Julian McAuley}.} \bibinfo{year}{2019}\natexlab{}.
\newblock \showarticletitle{Justifying Recommendations using Distantly-Labeled
  Reviews and Fine-Grained Aspects}. In \bibinfo{booktitle}{\emph{Proceedings
  of the 2019 Conference on Empirical Methods in Natural Language Processing
  and the 9th International Joint Conference on Natural Language Processing}}
  \emph{(\bibinfo{series}{EMNLP-IJCNLP '19})}. \bibinfo{pages}{188--197}.
\newblock


\bibitem[\protect\citeauthoryear{Ogawa, Nishikawa, Tokunaga, and Yokono}{Ogawa
  et~al\mbox{.}}{2020}]%
        {Ogawa:2020:LREC}
\bibfield{author}{\bibinfo{person}{Haruna Ogawa}, \bibinfo{person}{Hitoshi
  Nishikawa}, \bibinfo{person}{Takenobu Tokunaga}, {and}
  \bibinfo{person}{Hikaru Yokono}.} \bibinfo{year}{2020}\natexlab{}.
\newblock \showarticletitle{Gamification Platform for Collecting Task-oriented
  Dialogue Data}. In \bibinfo{booktitle}{\emph{Proceedings of the Twelfth
  Language Resources and Evaluation Conference}} \emph{(\bibinfo{series}{LREC
  '20})}. \bibinfo{pages}{7084--7093}.
\newblock


\bibitem[\protect\citeauthoryear{OpenAI}{OpenAI}{2023}]%
        {Openai:2023:arXiv}
\bibfield{author}{\bibinfo{person}{OpenAI}.} \bibinfo{year}{2023}\natexlab{}.
\newblock \bibinfo{title}{GPT-4 Technical Report}.
\newblock
\newblock
\showeprint[arxiv]{2303.08774}~[cs.CL]


\bibitem[\protect\citeauthoryear{Papenmeier, Frummet, and Kern}{Papenmeier
  et~al\mbox{.}}{2022}]%
        {Papenmeier:2022:CHIIR}
\bibfield{author}{\bibinfo{person}{Andrea Papenmeier},
  \bibinfo{person}{Alexander Frummet}, {and} \bibinfo{person}{Dagmar Kern}.}
  \bibinfo{year}{2022}\natexlab{}.
\newblock \showarticletitle{“Mhm...” – Conversational Strategies For
  Product Search Assistants}. In \bibinfo{booktitle}{\emph{ACM SIGIR Conference
  on Human Information Interaction and Retrieval}}
  \emph{(\bibinfo{series}{CHIIR '22})}. \bibinfo{pages}{36--46}.
\newblock


\bibitem[\protect\citeauthoryear{Petroni, Piktus, Fan, Lewis, Yazdani, De~Cao,
  Thorne, Jernite, Karpukhin, Maillard, Plachouras, Rockt{\"a}schel, and
  Riedel}{Petroni et~al\mbox{.}}{2021}]%
        {Petroni:2021:NAACL}
\bibfield{author}{\bibinfo{person}{Fabio Petroni}, \bibinfo{person}{Aleksandra
  Piktus}, \bibinfo{person}{Angela Fan}, \bibinfo{person}{Patrick Lewis},
  \bibinfo{person}{Majid Yazdani}, \bibinfo{person}{Nicola De~Cao},
  \bibinfo{person}{James Thorne}, \bibinfo{person}{Yacine Jernite},
  \bibinfo{person}{Vladimir Karpukhin}, \bibinfo{person}{Jean Maillard},
  \bibinfo{person}{Vassilis Plachouras}, \bibinfo{person}{Tim Rockt{\"a}schel},
  {and} \bibinfo{person}{Sebastian Riedel}.} \bibinfo{year}{2021}\natexlab{}.
\newblock \showarticletitle{{KILT}: a Benchmark for Knowledge Intensive
  Language Tasks}. In \bibinfo{booktitle}{\emph{Proceedings of the 2021
  Conference of the North American Chapter of the Association for Computational
  Linguistics: Human Language Technologies}} \emph{(\bibinfo{series}{NAACL
  '21})}. \bibinfo{pages}{2523--2544}.
\newblock


\bibitem[\protect\citeauthoryear{Qu, Yang, Croft, Zhang, Trippas, and Qiu}{Qu
  et~al\mbox{.}}{2019}]%
        {Qu:2019:CHIIR}
\bibfield{author}{\bibinfo{person}{Chen Qu}, \bibinfo{person}{Liu Yang},
  \bibinfo{person}{W.~Bruce Croft}, \bibinfo{person}{Yongfeng Zhang},
  \bibinfo{person}{Johanne~R. Trippas}, {and} \bibinfo{person}{Minghui Qiu}.}
  \bibinfo{year}{2019}\natexlab{}.
\newblock \showarticletitle{User Intent Prediction in Information-Seeking
  Conversations}. In \bibinfo{booktitle}{\emph{Proceedings of the 2019
  Conference on Human Information Interaction and Retrieval}}
  \emph{(\bibinfo{series}{CHIIR '19})}. \bibinfo{pages}{25--33}.
\newblock


\bibitem[\protect\citeauthoryear{Radlinski, Balog, Byrne, and
  Krishnamoorthi}{Radlinski et~al\mbox{.}}{2019}]%
        {Radlinski:2019:SIGDIAL}
\bibfield{author}{\bibinfo{person}{Filip Radlinski}, \bibinfo{person}{Krisztian
  Balog}, \bibinfo{person}{Bill Byrne}, {and} \bibinfo{person}{Karthik
  Krishnamoorthi}.} \bibinfo{year}{2019}\natexlab{}.
\newblock \showarticletitle{Coached Conversational Preference Elicitation: {A}
  Case Study in Understanding Movie Preferences}. In
  \bibinfo{booktitle}{\emph{Proceedings of the 20th Annual SIGdial Meeting on
  Discourse and Dialogue}} \emph{(\bibinfo{series}{SIGDIAL '19})}.
  \bibinfo{pages}{353--360}.
\newblock


\bibitem[\protect\citeauthoryear{Rajpurkar, Jia, and Liang}{Rajpurkar
  et~al\mbox{.}}{2018}]%
        {Rajpurkar:2018:ACL}
\bibfield{author}{\bibinfo{person}{Pranav Rajpurkar}, \bibinfo{person}{Robin
  Jia}, {and} \bibinfo{person}{Percy Liang}.} \bibinfo{year}{2018}\natexlab{}.
\newblock \showarticletitle{Know What You Don't Know: Unanswerable Questions
  for SQuAD}. In \bibinfo{booktitle}{\emph{Proceedings of the 56th Annual
  Meeting of the Association for Computational Linguistics (Volume 2: Short
  Papers)}} \emph{(\bibinfo{series}{ACL '18})}. \bibinfo{pages}{784--789}.
\newblock


\bibitem[\protect\citeauthoryear{Russell-Rose and Tate}{Russell-Rose and
  Tate}{2013}]%
        {RusselRose:2013:book}
\bibfield{author}{\bibinfo{person}{Tony Russell-Rose} {and}
  \bibinfo{person}{Tyler Tate}.} \bibinfo{year}{2013}\natexlab{}.
\newblock \bibinfo{booktitle}{\emph{Designing the Search Experience}}.
\newblock \bibinfo{publisher}{Morgan Kaufmann}.
\newblock


\bibitem[\protect\citeauthoryear{Serban, Lowe, Henderson, Charlin, and
  Pineau}{Serban et~al\mbox{.}}{2018}]%
        {Serban:2018:DD}
\bibfield{author}{\bibinfo{person}{Iulian~Vlad Serban}, \bibinfo{person}{Ryan
  Lowe}, \bibinfo{person}{Peter Henderson}, \bibinfo{person}{Laurent Charlin},
  {and} \bibinfo{person}{Joelle Pineau}.} \bibinfo{year}{2018}\natexlab{}.
\newblock \showarticletitle{A Survey of Available Corpora For Building
  Data-Driven Dialogue Systems: The Journal Version}.
\newblock \bibinfo{journal}{\emph{Dialogue \& Discourse}} \bibinfo{volume}{9},
  \bibinfo{number}{1} (\bibinfo{year}{2018}), \bibinfo{pages}{1--49}.
\newblock


\bibitem[\protect\citeauthoryear{Szpektor, Cohen, Elidan, Fink, Hassidim,
  Keller, Kulkarni, Ofek, Pudinsky, Revach, Salant, and Matias}{Szpektor
  et~al\mbox{.}}{2020}]%
        {Szpektor:2020:WWW}
\bibfield{author}{\bibinfo{person}{Idan Szpektor}, \bibinfo{person}{Deborah
  Cohen}, \bibinfo{person}{Gal Elidan}, \bibinfo{person}{Michael Fink},
  \bibinfo{person}{Avinatan Hassidim}, \bibinfo{person}{Orgad Keller},
  \bibinfo{person}{Sayali Kulkarni}, \bibinfo{person}{Eran Ofek},
  \bibinfo{person}{Sagie Pudinsky}, \bibinfo{person}{Asaf Revach},
  \bibinfo{person}{Shimi Salant}, {and} \bibinfo{person}{Yossi Matias}.}
  \bibinfo{year}{2020}\natexlab{}.
\newblock \showarticletitle{Dynamic Composition for Conversational Domain
  Exploration}. In \bibinfo{booktitle}{\emph{Proceedings of The Web Conference
  2020}} \emph{(\bibinfo{series}{WWW '20})}. \bibinfo{pages}{872--883}.
\newblock


\bibitem[\protect\citeauthoryear{Thomas, McDuff, Czerwinski, and
  Craswell}{Thomas et~al\mbox{.}}{2017}]%
        {Thomas:2017:CAIR}
\bibfield{author}{\bibinfo{person}{Paul Thomas}, \bibinfo{person}{Daniel
  McDuff}, \bibinfo{person}{Mary Czerwinski}, {and} \bibinfo{person}{Nick
  Craswell}.} \bibinfo{year}{2017}\natexlab{}.
\newblock \showarticletitle{MISC: A data set of information-seeking
  conversations}. In \bibinfo{booktitle}{\emph{Proceedings of the 1st
  International Workshop on Conversational Approaches to Information
  Retrieval}} \emph{(\bibinfo{series}{CAIR '17})}.
\newblock


\bibitem[\protect\citeauthoryear{Voicebot.ai}{Voicebot.ai}{2020}]%
        {Voicebot:2022:website}
\bibfield{author}{\bibinfo{person}{Business~Wire Voicebot.ai}.}
  \bibinfo{year}{2020}\natexlab{}.
\newblock \bibinfo{title}{Number of digital voice assistants in use worldwide
  from 2019 to 2024 (in billions)*}.
\newblock
\newblock
\urldef\tempurl%
\url{https://www.statista.com/statistics/973815/worldwide-digital-voice-assistant-in-use/}
\showURL{%
\tempurl}
\newblock
\shownote{Accessed: 2022-02-16.}


\bibitem[\protect\citeauthoryear{Yu, Liu, Yang, Xiong, Bennett, Gao, and
  Liu}{Yu et~al\mbox{.}}{2020}]%
        {Yu:2020:SIGIR}
\bibfield{author}{\bibinfo{person}{Shi Yu}, \bibinfo{person}{Jiahua Liu},
  \bibinfo{person}{Jingqin Yang}, \bibinfo{person}{Chenyan Xiong},
  \bibinfo{person}{Paul Bennett}, \bibinfo{person}{Jianfeng Gao}, {and}
  \bibinfo{person}{Zhiyuan Liu}.} \bibinfo{year}{2020}\natexlab{}.
\newblock \showarticletitle{Few-Shot Generative Conversational Query
  Rewriting}. In \bibinfo{booktitle}{\emph{Proceedings of the 43rd
  International ACM SIGIR Conference on Research and Development in Information
  Retrieval}} \emph{(\bibinfo{series}{SIGIR '20})}.
  \bibinfo{pages}{1933--1936}.
\newblock


\bibitem[\protect\citeauthoryear{Zaib, Zhang, Sheng, Mahmood, and Zhang}{Zaib
  et~al\mbox{.}}{2022}]%
        {Zaib:2022:KAIS}
\bibfield{author}{\bibinfo{person}{Munazza Zaib}, \bibinfo{person}{Wei~Emma
  Zhang}, \bibinfo{person}{Quan~Z. Sheng}, \bibinfo{person}{Adnan Mahmood},
  {and} \bibinfo{person}{Yang Zhang}.} \bibinfo{year}{2022}\natexlab{}.
\newblock \showarticletitle{Conversational question answering: a survey}.
\newblock \bibinfo{journal}{\emph{Knowledge and Information Systems}}
  \bibinfo{volume}{64}, \bibinfo{number}{12} (\bibinfo{year}{2022}),
  \bibinfo{pages}{3151--3195}.
\newblock


\bibitem[\protect\citeauthoryear{Zamani, Trippas, Dalton, and Radlinski}{Zamani
  et~al\mbox{.}}{2022}]%
        {Zamani:2022:arXiv}
\bibfield{author}{\bibinfo{person}{Hamed Zamani}, \bibinfo{person}{Johanne~R.
  Trippas}, \bibinfo{person}{Jeff Dalton}, {and} \bibinfo{person}{Filip
  Radlinski}.} \bibinfo{year}{2022}\natexlab{}.
\newblock \bibinfo{title}{Conversational Information Seeking}.
\newblock
\newblock
\showeprint[arxiv]{2201.08808}~[cs.IR]


\bibitem[\protect\citeauthoryear{Zhang, Takanobu, Zhu, Huang, and Zhu}{Zhang
  et~al\mbox{.}}{2020}]%
        {Zhang:2020:SciChinaTechSci}
\bibfield{author}{\bibinfo{person}{Zheng Zhang}, \bibinfo{person}{Ryuichi
  Takanobu}, \bibinfo{person}{Qi Zhu}, \bibinfo{person}{MinLie Huang}, {and}
  \bibinfo{person}{XiaoYan Zhu}.} \bibinfo{year}{2020}\natexlab{}.
\newblock \showarticletitle{Recent advances and challenges in task-oriented
  dialog systems}.
\newblock \bibinfo{journal}{\emph{Science China Technological Sciences}}
  \bibinfo{volume}{63}, \bibinfo{number}{10} (\bibinfo{year}{2020}),
  \bibinfo{pages}{2011--2027}.
\newblock


\bibitem[\protect\citeauthoryear{Zhou, Zhou, Zhao, Wang, and Wen}{Zhou
  et~al\mbox{.}}{2020}]%
        {Zhou:2020:COLING}
\bibfield{author}{\bibinfo{person}{Kun Zhou}, \bibinfo{person}{Yuanhang Zhou},
  \bibinfo{person}{Wayne~Xin Zhao}, \bibinfo{person}{Xiaoke Wang}, {and}
  \bibinfo{person}{Ji-Rong Wen}.} \bibinfo{year}{2020}\natexlab{}.
\newblock \showarticletitle{Towards Topic-Guided Conversational Recommender
  System}. In \bibinfo{booktitle}{\emph{Proceedings of the 28th International
  Conference on Computational Linguistics}} \emph{(\bibinfo{series}{COLING
  '20})}. \bibinfo{pages}{4128--4139}.
\newblock


\bibitem[\protect\citeauthoryear{Zuo, Hu, Yu, Li, Zhao, and Joe-Wong}{Zuo
  et~al\mbox{.}}{2022}]%
        {Zuo:2022:CIKM}
\bibfield{author}{\bibinfo{person}{Jinhang Zuo}, \bibinfo{person}{Songwen Hu},
  \bibinfo{person}{Tong Yu}, \bibinfo{person}{Shuai Li},
  \bibinfo{person}{Handong Zhao}, {and} \bibinfo{person}{Carlee Joe-Wong}.}
  \bibinfo{year}{2022}\natexlab{}.
\newblock \showarticletitle{Hierarchical Conversational Preference Elicitation
  with Bandit Feedback}. In \bibinfo{booktitle}{\emph{Proceedings of the 31st
  ACM International Conference on Information \& Knowledge Management}}
  \emph{(\bibinfo{series}{CIKM '22})}. \bibinfo{pages}{2827--2836}.
\newblock


\end{thebibliography}
